# Supergaussian pulse distortion in singlemode fibers with arbitrary dispersion

JOSÉ CAPMANY*

*ITEAM Research Institute, Universitat Politècnica de València, Camino de Vera s/n, 46022 Valencia, Spain*
*jcapmany@iteam.upv.es

**Abstract:** Marcuse's theory of pulse distortion in single-mode fibers provides closed-form results for Gaussian pulses launched from sources of arbitrary spectral width. We extend the theory to chirped super-Gaussian pulses of arbitrary order, single- or multi-line sources and an arbitrary number of dispersion coefficients. The rms output width is obtained exactly from the moments of the Wigner distribution, and all the pulse moments involved reduce to finite sums of Gamma functions. Super-Gaussian pulses are considerably more sensitive than Gaussian pulses to higher-order dispersion and are compressed less efficiently by chirp. Near the zero-dispersion wavelength their rms width is governed by weak spectral side lobes and overestimates the broadening of the pulse energy. For pulse sequences, interference between neighboring pulses leaves all time moments unchanged when the input pulses do not overlap, but it redistributes energy locally with a visibility set by the source spectrum filtered by the pulse spectrum.

## 1. Introduction

In a series of three papers, Marcuse developed a theory of pulse distortion in single-mode fibers that accounts simultaneously for fiber dispersion and for the finite spectral width of the source [1–3]. Part 1 [1] considered a Gaussian power-modulated pulse from a source with a Gaussian spectrum of arbitrary width, expanded the propagation constant to third order in frequency, and derived a closed-form spectral function, an exact Gaussian solution in the absence of second-order dispersion, and the rms width

$$\frac{\sigma^2}{\sigma_0^2} = 1 + 4D^2(1 + V^2) + 9B^2(1 + V^2)^2, \tag{1}$$

with $V = WT$, $D = \ddot{\beta}_0 z/(2T^2)$ and $B = \dddot{\beta}_0 z/(6T^3)$. Part 2 [2] treated sources emitting several discrete lines, and Part 3 [3] added a linear frequency chirp $C = T\Delta\omega$, obtaining

$$\frac{\sigma^2}{\sigma_0^2} = (1 + 2DC)^2 + 4D^2(1 + V^2) + 9B^2(1 + V^2 + C^2)^2. \tag{2}$$

Gaussian pulses with second-order dispersion and finite source bandwidth were also analyzed by Miyagi and Nishida [4,5], and the theory was later extended to Gaussian pulses in fibers and waveguides with an arbitrary number of dispersion coefficients, including source chirp and linewidth [6]. Amemiya showed that dispersion described by the coefficient $\beta_N$ limits the transmission length in proportion to $1/B_0^N$, $B_0$ being the bit rate [7], and a general description of spectrally partially coherent pulses was given by Lajunen et al. [8].

All of these results rely on the Gaussian shape of the input pulse. Pulses produced by external modulators and directly modulated lasers, the individual bits of non-return-to-zero signals, and the flat-top pulses synthesized by pulse shapers have flatter tops and steeper edges, and are commonly modeled by super-Gaussian functions [9,10]. Agrawal and Potasek [10] introduced a super-Gaussian model with edge-concentrated chirp for directly modulated semiconductor lasers and analyzed it numerically. Using the moment method of Anderson and Askne [11], Anderson and Lisak [12] then showed that for a coherent source the rms width varies parabolically with distance irrespective of pulse form or chirp. They noted that this holds for any dispersion order, evaluated the coefficients in closed form for first-order dispersion, and found the Gaussian shape superior; pulse broadening in dispersive fibers was analyzed further in [13], and the effect of higher-order dispersion on the broadening of partially incoherent light was studied in [14]. To our knowledge, the combined effect of pulse shape, source spectral width and higher-order dispersion, which is the situation considered by Marcuse, has not been treated analytically.

Here we extend Marcuse's theory to chirped super-Gaussian pulses of arbitrary order $m$, single- or multi-line Gaussian sources and an arbitrary number of dispersion coefficients. The model is described in Section 2, and an exact moment theorem for the rms width is derived in Section 3. The pulse moments required are obtained in Section 4, together with their asymptotic form for large $m$. Section 5 presents the rms width for arbitrary dispersion, with an explicit expression for Gaussian pulses and a discussion of first- and second-order dispersion, higher orders and multi-line sources. Optimum input widths and chirp compression are considered in Section 6, where the results of Anderson and Lisak are recovered. Section 7 deals with pulse shapes and with the interpretation of the rms width for steep-edged pulses. Section 8 extends the formalism to pulse sequences, analyzes the interference between neighboring pulses, and presents eye diagrams and dispersion-limited capacity curves. Section 9 gives numerical examples for standard fiber, and Section 10 compares the results with previous work. The derivations are given in the Appendices and the numerical checks in Supplement 1.

## 2. Model

We follow Marcuse's formulation [1]. A source with carrier frequency $\omega_0$ and stationary random fluctuations, described by the normalized power spectrum $S(\omega')$ of the frequency offset $\omega'$, is modulated by a deterministic complex envelope $u(t)$. The ensemble-averaged output power is then the sum of the powers carried by the individual spectral components of the source (Appendix A),

$$P(z,t) = \int S(\omega')|a(z,t;\omega')|^2 d\omega',$$
$$a(z,t;\omega') = \frac{1}{2\pi}\int \tilde{u}(\omega-\omega')\, e^{-i\Delta\beta(\omega)z} e^{i\omega t} d\omega, \tag{3}$$

where $\tilde{u}$ is the Fourier transform of $u$, time is measured in the frame moving at the group velocity at $\omega_0$, and

$$\Delta\beta(\omega) = \beta(\omega_0+\omega) - \beta_0 - \beta_1\omega = \sum_{n\geq 2} \frac{\beta_n}{n!}\omega^n. \tag{4}$$

In Marcuse's notation $\beta_2 = \ddot{\beta}_0$ and $\beta_3 = \dddot{\beta}_0$. The source spectrum is a single Gaussian line, as in [1,3], or a set of Gaussian lines with weights $p_j$ and offsets $\Omega_j$, as in [2],

$$S(\omega') = \sum_j \frac{p_j}{\sqrt{\pi}W}\exp\left[-\frac{(\omega'-\Omega_j)^2}{W^2}\right], \qquad \sum_j p_j = 1, \tag{5}$$

with $V = WT$ as before. The input envelope is a chirped super-Gaussian of order $m$,

$$u(t) = \exp\left[-\frac{1-iC}{2}\left(\frac{t}{T}\right)^{2m}\right]. \tag{6}$$

The phase of Eq. (6) follows the envelope, as in the model of Agrawal and Potasek [10], and the instantaneous frequency $\omega_i = mC\, t^{2m-1}/T^{2m}$ is therefore concentrated at the pulse edges for $m > 1$. For $m = 1$ it reduces to Marcuse's linear chirp, $\omega_i = Ct/T^2$, which sweeps by $\Delta\omega = C/T$ over the half-width $T$.[1] For $m = 1$ and $C = 0$, Eq. (6) is Marcuse's pulse, with input power $e^{-t^2/T^2}$ and rms width $\sigma_0 = T/\sqrt{2}$; as $m \to \infty$ it tends to a rectangular pulse of duration $2T$. We keep Marcuse's normalized parameters and generalize them to any order as

[1] For $m = 1$, $C$ is the linear chirp parameter of Marcuse [3]. For $m \neq 1$, $C$ has to be understood as the parameter $\alpha$ of the model of Agrawal and Potasek [10], which plays the role of a linewidth enhancement factor of a directly modulated laser.

$$D_n = \frac{\beta_n z}{n!\,T^n}, \tag{7}$$

so that $D_2 \equiv D$ and $D_3 \equiv B$; we use $D$ and $B$ whenever only these two terms are present. Following the usual convention, $\beta_2$ ($D$) describes first-order dispersion, $\beta_3$ ($B$) second-order dispersion, and in general $\beta_n$ ($D_n$) describes dispersion of order $n-1$. The spectral phase in Eq. (3) is then $\Delta\beta z = \sum_n D_n\,(\omega T)^n$. For $m > 1$ the transform $\tilde{u}$ cannot be written in elementary form and the pulse shapes have to be computed numerically. The rms width, on the other hand, can be obtained exactly, as shown next.

## 3. Moment theorem for the rms width

Consider a single source component $\omega'$ and write the output spectrum as $A_z(\omega) = \tilde{u}(\omega - \omega')e^{-i\Delta\beta(\omega)z} = |A_z|e^{i\psi_z}$, normalized to unit energy. Multiplication by $t$ corresponds to $i\,\partial_\omega$ in the frequency domain, so

$$\langle t\rangle_z = \int (-\psi_z{}')|A_z|^2 \frac{d\omega}{2\pi},$$

$$\langle t^2\rangle_z = \int [(\partial_\omega|A_z|)^2 + \psi_z{}'^2|A_z|^2]\frac{d\omega}{2\pi}. \tag{8}$$

Dispersion changes only the spectral phase, $\psi_z = \psi_0 - \Delta\beta\, z$, so $-\psi_z{}' = \tau_0(\omega) + z\,\tau(\omega)$, where $\tau_0 = -\psi_0{}'$ is the spectral group delay of the input pulse and

$$\tau(\omega) = \frac{d\Delta\beta}{d\omega} = \sum_{n\geq 2} \frac{\beta_n}{(n-1)!}\omega^{n-1} \tag{9}$$

is the group delay per unit length relative to the carrier. Hence $\langle t\rangle_z = \langle\tau_0\rangle + z\langle\tau\rangle$ and $\langle t^2\rangle_z = \langle t^2\rangle_0 + 2z\langle\tau_0\tau\rangle + z^2\langle\tau^2\rangle$, with spectral averages over $|A_0|^2$. The averages involving $\tau_0$ are conveniently written with the Wigner distribution of the input [15],

$$\mathcal{W}(t,\Omega) = \frac{1}{2\pi}\int u\left(t+\frac{s}{2}\right)u^*\left(t-\frac{s}{2}\right)e^{-i\Omega s}ds, \tag{10}$$

whose conditional mean in time at fixed frequency is the spectral group delay: $\iint t\, g(\Omega)\mathcal{W}\,dt\,d\Omega = \langle\tau_0 g\rangle$ for any function $g$. A source component $\omega'$ shifts $\mathcal{W}$ in frequency. Averaging over $S(\omega')$, with each component carrying the same energy, and subtracting the squared mean yields

$$\sigma^2(z) = \sigma_0^2 + 2z\,\mathrm{Cov}[t,\tau(X)] + z^2\,\mathrm{Var}[\tau(X)], \qquad X = \Omega + \omega'. \tag{11}$$

The averages are taken over $\mathcal{W}(t,\Omega)S(\omega')$, and $X$ is the frequency at which a given portion of the pulse travels; the mean output delay is $\langle t\rangle_z = z\langle\tau(X)\rangle$. Equation (11) is exact for any dispersion relation, any chirp and any source spectrum, provided the moments involved are finite. It displays the parabolic dependence on distance found by Anderson and Lisak [12] and provides both coefficients for partially coherent sources and arbitrary dispersion. Although the Wigner distribution may take negative values, only its polynomial moments are needed, and these are ordinary Weyl-ordered expectation values.

Since $\sigma$ is the exact standard deviation of the output power, Chebyshev's inequality ensures that at most a fraction $1/k^2$ of the energy lies outside $|t - \langle t\rangle| > k\sigma$. The rms width thus provides a rigorous bound on the spreading of the pulse energy for any pulse shape. As shown in Section 7, this bound can be rather loose for pulses with steep edges.

## 4. Moments of the super-Gaussian pulse

With a polynomial $\tau(X)$, Eq. (11) requires the joint moments of $(t,\Omega)$ over $\mathcal{W}$. We measure time in units of $T$ and frequency in units of $1/T$, and denote normalized quantities with hats. The derivatives of the envelope of Eq. (6) have the form $u^{(k)} = P_k(\hat{t})\,u$, with complex polynomials generated by

$$P_0 = 1, \qquad P_{k+1} = P_k{}' - m(1-iC)\,\hat{t}^{\,2m-1}P_k. \tag{12}$$

Integrating Eq. (10) over frequency against $\Omega^l$ gives the Weyl-ordered joint moments (Appendix B)

$$\langle \hat{t}^{\,j}\hat{\Omega}^l\rangle_W = \mathrm{Re}\left[\left(\frac{-i}{2}\right)^l \sum_{k=0}^{l}\binom{l}{k}(-1)^{l-k}\,\mathcal{J}[\hat{t}^{\,j}P_kP_{l-k}^*]\right],$$

$$\mathcal{J}[\hat{t}^{\,p}] = \frac{\Gamma\left(\frac{p+1}{2m}\right)}{\Gamma\left(\frac{1}{2m}\right)}\ (p\ \text{even}), \qquad \mathcal{J}[\hat{t}^{\,p}] = 0\ (p\ \text{odd}), \tag{13}$$

where $\mathcal{J}$ acts term by term on the polynomial; it is the normalized average over $|u|^2 = e^{-\hat{t}^{2m}}$, which is independent of the chirp. All the joint moments are therefore finite sums of Gamma functions and remain finite for any finite $m$. For the unchirped pulse the moments vanish unless $j$ and $l$ are both even; for $j = 0$, Eq. (13) reduces to the spectral moments $\hat{\mu}_{2k} = \mathcal{J}[|P_k|^2]$. With $a = 1/(2m)$, the lowest-order constants of the unchirped pulse are

$$\hat{\sigma}_0^2 = \frac{\Gamma(3a)}{\Gamma(a)}, \qquad \hat{\mu}_2 = \frac{(2m-1)\,\Gamma(1-a)}{4\,\Gamma(1+a)}, \tag{14}$$

$$\hat{\mu}_4 = \frac{(2m-1)(2m-3)(4m-1)\,\Gamma(1-3a)}{16\,\Gamma(1+a)}. \tag{15}$$

Table 1 lists these constants. The Gaussian values are $\hat{\sigma}_0^2 = \hat{\mu}_2 = 1/2$ and $\hat{\mu}_4 = 3/4$. For large $m$ the moments are dominated by the two edges, whose profile near $|\hat{t}| = 1$ tends to $f(s) = \exp(-e^s/2)$ with $s = 2m(|\hat{t}| - 1)$, and one finds (Appendix C)

$$\hat{\sigma}_0^2 \to \frac{1}{3}, \qquad \hat{\mu}_{2k} \simeq \frac{\mathcal{T}_k}{2}\, m^{2k-1}, \qquad \mathcal{T}_k = 1,\ 2,\ 16,\ 272,\ 7936,\ \ldots, \tag{16}$$

where $\mathcal{T}_k$ are the tangent numbers. Thus $\hat{\mu}_2 \simeq m/2$, $\hat{\mu}_4 \simeq m^3$, $\hat{\mu}_6 \simeq 8m^5$, and all spectral moments diverge in the rectangular limit, whose spectrum decays only as $\Omega^{-2}$. The convergence to Eq. (16) is slow for higher $k$; at $m = 32$ the ratio of $\hat{\mu}_{2k}$ to its asymptote is 1.002, 0.967 and 0.913 for $k = 1$, 2 and 3. The origin of this behavior can be seen in Fig. 1: the steep edges of the pulse give rise to spectral side lobes that decay much more slowly than the Gaussian spectrum.

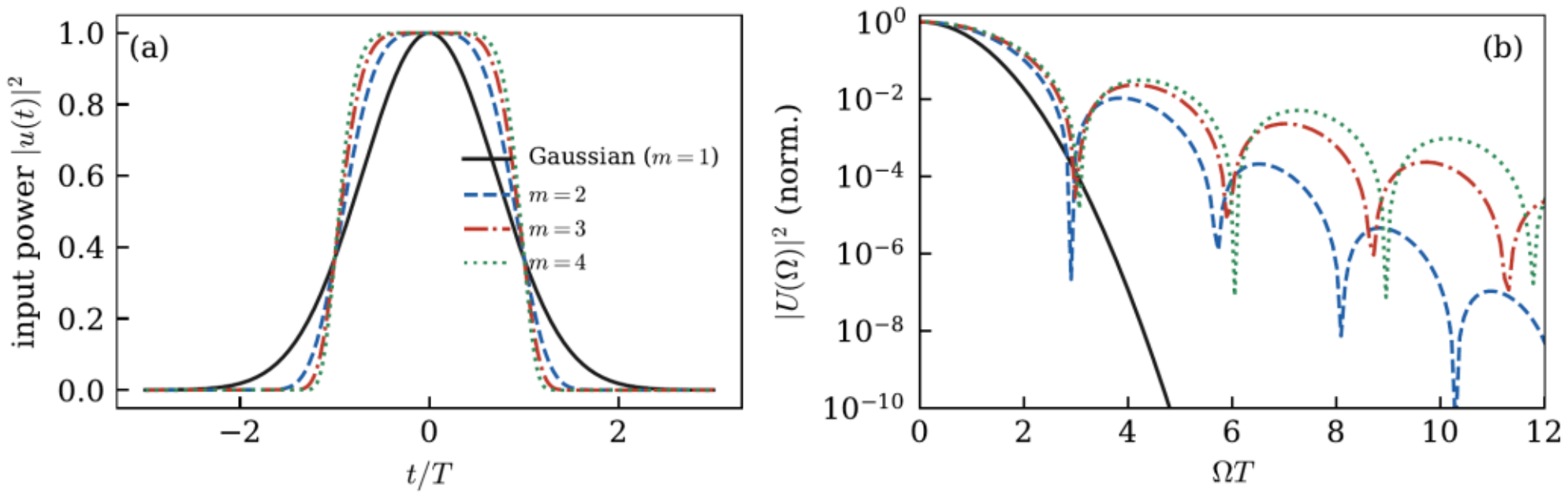

**Fig. 1.** Super-Gaussian input pulses and their spectra. (a) Input power $|u(t)|^2$ versus normalized time $t/T$ for $m = 1$ (Gaussian, solid), 2 (dashed), 3 (dash-dotted) and 4 (dotted), all with the same $1/e$ half-width $T$ and no chirp. (b) Corresponding normalized power spectra $|\tilde{u}(\Omega)|^2$ on a logarithmic scale versus $\Omega T$. The slowly decaying side lobes of the steep-edged pulses carry little energy but dominate the higher spectral moments.

**Table 1. Shape Constants of the Unchirped Super-Gaussian Pulse[a]**

| $m$ | $\hat{\sigma}_0^2$ | $\hat{\mu}_2$ | $\hat{\mu}_4$ | $4\hat{\sigma}_0^2\hat{\mu}_2$ | $K_3/\hat{\sigma}_0^2$ |
|---|---|---|---|---|---|
| 1 | 0.5000 | 0.5000 | 0.750 | 1.000 | 1.00 |
| 2 | 0.3380 | 1.0140 | 5.250 | 1.371 | 12.49 |
| 3 | 0.3184 | 1.5209 | 19.703 | 1.937 | 54.61 |
| 4 | 0.3146 | 2.0249 | 49.982 | 2.548 | 145.83 |
| 6 | 0.3153 | 3.0291 | 181.983 | 3.820 | 548.10 |
| 8 | 0.3175 | 4.0314 | 449.228 | 5.120 | 1363.76 |

| $m$ | $\hat{\sigma}_0^2$ | $\hat{\mu}_2$ | $\hat{\mu}_4$ | $4\hat{\sigma}_0^2\hat{\mu}_2$ | $K_3/\hat{\sigma}_0^2$ |
|---|---|---|---|---|---|
| $m \gg 1$ | $1/3$ | $m/2$ | $m^3$ | $2m/3$ | $3m^3$ |

[a]Equations (14) and (15); the last row gives the large-$m$ behavior of Eq. (16). $4\hat{\sigma}_0^2\hat{\mu}_2$ is the rms time–bandwidth product normalized to its Gaussian value and governs chirp compression and the optimum input width (Section 6). $K_3 = \hat{\mu}_4 - \hat{\mu}_2^2$, and $K_3/\hat{\sigma}_0^2$ is the coefficient of $9B^2$ in $(\sigma/\sigma_0)^2$ for $V = C = 0$.

## 5. rms width for arbitrary dispersion

### 5.1 General result

With Eqs. (7) and (9), $z\tau(X)/T = \sum_n n\, D_n \hat{X}^{n-1}$. Inserting this into Eq. (11) gives the rms width for any number of dispersion coefficients:

$$\frac{\sigma^2}{\sigma_0^2} = 1 + \frac{2}{\hat{\sigma}_0^2}\sum_n n\, D_n\, \mathrm{Cov}\big[\hat{t}, \hat{X}^{n-1}\big] + \frac{1}{\hat{\sigma}_0^2}\sum_{n,k} n\, k\, D_n D_k\big(\langle \hat{X}^{n+k-2}\rangle - \langle \hat{X}^{n-1}\rangle\langle \hat{X}^{k-1}\rangle\big), \tag{17}$$

with mean delay $\langle t\rangle_z/T = \sum_n n\, D_n \langle \hat{X}^{n-1}\rangle$. Because $\hat{\omega}'$ is independent of $(\hat{t}, \hat{\Omega})$, the moments of $\hat{X} = \hat{\Omega} + \hat{\omega}'$ follow from Eq. (13) through the binomial expansion

$$\langle \hat{t}^{\,j}\hat{X}^n\rangle = \sum_{a=0}^{n}\binom{n}{a}\langle \hat{t}^{\,j}\hat{\Omega}^a\rangle_W \langle \hat{\omega}'^{\,n-a}\rangle,$$

$$\langle \hat{\omega}'^{\,c}\rangle = \sum_j p_j \sum_{i\ \mathrm{even}}\binom{c}{i}\hat{\Omega}_j^{c-i}\,(i-1)!!\left(\frac{V^2}{2}\right)^{i/2}. \tag{18}$$

Equations (13), (17) and (18) express the rms width of a chirped super-Gaussian pulse of any order, with a single- or multi-line Gaussian source, for any number of dispersion coefficients, in terms of Gamma functions only. Because the envelope and the phase of Eq. (6) are even in $t$, a single-line source leads to three selection rules. Only coefficients with even $n$ ($\beta_2, \beta_4, \ldots$) contribute to the term linear in $z$, and only in the presence of chirp. Only coefficients with odd $n$ ($\beta_3, \beta_5, \ldots$) shift the mean delay. In the variance, each coefficient couples only to coefficients with $n$ of the same parity. The last two rules no longer hold for an asymmetric multi-line source (Section 5.5).

### 5.2 Gaussian pulses: explicit closed form

For $m = 1$ and a single-line source, $(\hat{t}, \hat{X})$ is jointly Gaussian with $\mathrm{Var}\,\hat{t} = 1/2$, $\mathrm{Var}\,\hat{X} = s^2 = (1 + C^2 + V^2)/2$ and $\mathrm{Cov}(\hat{t}, \hat{X}) = C/2$. Stein's lemma, $\langle \hat{t}\, g(\hat{X})\rangle = \mathrm{Cov}(\hat{t}, \hat{X})\langle g'(\hat{X})\rangle$, and the Gaussian moments $\langle \hat{X}^{2j}\rangle = (2j-1)!!\, s^{2j}$ then reduce Eq. (17) to

$$\frac{\sigma^2}{\sigma_0^2} = 1 + 2C\sum_{n\ \mathrm{even}} n\,(n-1)!!\, D_n\, s^{n-2} + 2\sum_{n+k\ \mathrm{even}} n\, k\, D_n D_k\big[(n+k-3)!! - \delta_{n,k}^{\mathrm{odd}}(n-2)!!\,(k-2)!!\big]s^{n+k-2}, \tag{19}$$

where $\delta_{n,k}^{\mathrm{odd}} = 1$ if $n$ and $k$ are both odd and 0 otherwise, and $(-1)!! = 1$. Equation (19) gives the rms width of Gaussian pulses with chirp, source width and any number of dispersion coefficients, the problem considered in [6], and reduces to Marcuse's Eq. (2) when only $D_2$ and $D_3$ are retained.

### 5.3 First- and second-order dispersion

For $D$ and $B$ only and a single-line source, Eq. (17) requires $\langle \hat{t}\hat{X}\rangle$, $\hat{M}_2 = \langle \hat{X}^2\rangle$ and $\hat{M}_4 = \langle \hat{X}^4\rangle$. For the chirped pulse of Eq. (6), Eq. (13) gives $\langle \hat{t}\hat{\Omega}\rangle_W = C/2$ for every $m$, and

$$\hat{M}_2 = (1 + C^2)\hat{\mu}_2 + \frac{V^2}{2}, \qquad \hat{M}_4 = (1 + C^2)\hat{\mu}_4\left[1 + \frac{(12m-9)C^2}{4m-1}\right] + 3(1 + C^2)\hat{\mu}_2 V^2 + \frac{3}{4}V^4, \tag{20}$$

and the rms width becomes

$$\frac{\sigma^2}{\sigma_0^2} = 1 + \frac{2DC + 4D^2\hat{M}_2 + 9B^2(\hat{M}_4 - \hat{M}_2^2)}{\hat{\sigma}_0^2}. \quad (21)$$

With the Gaussian constants, $\hat{M}_2 = (1 + V^2 + C^2)/2$ and $\hat{M}_4 = 3(1 + V^2 + C^2)^2/4$, and Eq. (21) reduces to Marcuse's Eq. (2). The following remarks can be made.

*(i) Narrow source at the zero-dispersion wavelength.* For $V = C = D = 0$ the coefficient of $9B^2$ is $K_3/\hat{\sigma}_0^2$. It is equal to 1 for the Gaussian pulse, but takes the values 12.5, 54.6 and 146 for $m = 2$, 3 and 4 (Table 1) and grows as $3m^3$ for large $m$.

*(ii) Broad source.* For $V \gg 1$, $\hat{M}_2 \to V^2/2$ and $\hat{M}_4 - \hat{M}_2^2 \to V^4/2$, independently of $m$. The absolute broadening $\sigma^2 - \sigma_0^2$ then becomes independent of the pulse shape, and Marcuse's conclusion that the rms width grows as $W$ away from, and as $W^2$ at, the zero-dispersion wavelength holds for all $m$.

*(iii) Chirp.* The linear term $2DC/\hat{\sigma}_0^2$ reduces to Marcuse's $4DC$ for $m = 1$ and increases as $1/\hat{\sigma}_0^2$ for $m > 1$, by factors of 1.48 and 1.57 for $m = 2$ and 3; this is the linear coefficient $A$ of Anderson and Lisak [12]. The pulse is initially compressed when $D$ and $C$ have opposite signs. The chirp also broadens the spectrum, through the factor $1 + C^2$ in $\hat{M}_2$ and $\hat{M}_4$, and the additional factor in square brackets in Eq. (20) couples chirp and second-order dispersion more strongly for steep-edged pulses.

*(iv) Source width versus chirp at $\lambda_0$.* For a Gaussian pulse at $D = 0$, $V^2$ and $C^2$ enter only through $1 + V^2 + C^2$, so a source width and an equal chirp give the same rms width [3]. According to Eq. (20) this equivalence is a property of the Gaussian shape only: $V^2/2$ adds to $\hat{\mu}_2$, whereas $C^2$ multiplies it, and the fourth-order terms differ as well. For $m = 3$, $D = 0$ and $B = 0.3$, $V = 1.5$ gives $\sigma^2/\sigma_0^2 = 69.1$, whereas $C = 1.5$ gives 1001; for $m = 1$ both give 9.56 (Fig. 7).

### *5.4 Higher-order dispersion*

Higher-order dispersion coefficients involve higher spectral moments. For $V = C = 0$ the variance term of $\beta_n$ alone is proportional to $K_n = \hat{\mu}_{2n-2} - \hat{\mu}_{n-1}^2$, which by Eq. (16) grows as $m^{2n-3}$. In addition, with chirp, third-order dispersion contributes to the term linear in $z$ in Eq. (17),

$$\left.\frac{\sigma^2}{\sigma_0^2}\right|_{\beta_4} = 1 + \frac{2D_4C}{\hat{\sigma}_0^2}\left[(4m-1)(1 + C^2)\hat{\mu}_2 + \frac{3}{2}V^2\right] + \frac{16D_4^2}{\hat{\sigma}_0^2}\langle\hat{X}^6\rangle, \quad (22)$$

so that a chirped pulse is initially compressed by third-order dispersion when $D_4C < 0$, even in the absence of $\beta_2$. As Fig. 2(a) shows, this compression is significant only for nearly Gaussian pulses. With $C = 1$ and $V = 0$, a Gaussian pulse is compressed to $0.84\,\sigma_0$ at $D_4 = -0.025$, whereas the minimum widths for $m = 2$, 3 and 4 are 0.95, 0.98 and $0.99\,\sigma_0$, since the quadratic term, which depends on $\hat{\mu}_6$, dominates almost from the start. The symmetric distortion produced by $\beta_4$ alone is shown in Fig. 2(b).

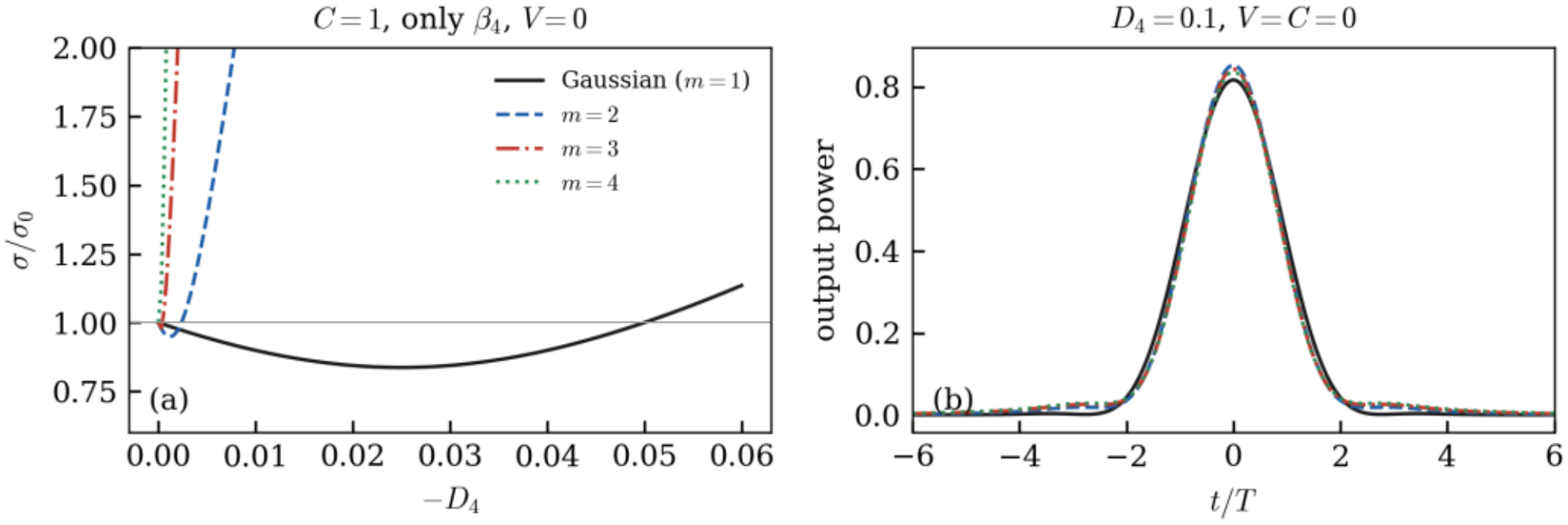

**Fig. 2.** Effect of third-order dispersion. (a) rms width $\sigma/\sigma_0$ of chirped pulses ($C = 1$, $V = 0$) in a fiber with only third-order dispersion, versus $-D_4$, from Eq. (22); the initial decrease is compression by $\beta_4$. (b) Output power (input peak = 1) versus $t/T$ for unchirped pulses with $D_4 = 0.1$ and $V = 0$, computed from Eq. (3). Line styles as in Fig. 1: $m = 1$ (solid), 2 (dashed), 3 (dash-dotted), 4 (dotted).

### *5.5 Multi-line sources*

For the multi-line spectrum of Eq. (5), Eq. (18) contains the moments of the line offsets $\hat{\Omega}_j$. A symmetric set of lines increases $\hat{M}_2$ like an additional source width, and modifies $\hat{M}_4$ according to the fourth moment of the line offsets. An asymmetric set has nonzero odd moments $\langle\hat{\omega}'\rangle$ and $\langle\hat{\omega}'^3\rangle$, which shift the mean delay through $\beta_2$ and, with both $D$ and $B$ present, add the cross term $12DB(\hat{M}_3 - \hat{M}_1\hat{M}_2)/\hat{\sigma}_0^2$ to Eq. (21). This coupling between first- and second-order dispersion through the asymmetry of the source generalizes, for the rms width, the asymmetry effects discussed by Marcuse in Part 2 [2] to arbitrary pulse order.

## 6. Optimum input width and chirp compression

### *6.1 Optimum input width*

When a single dispersion coefficient $\beta_n$ dominates and $V = C = 0$, Eq. (17) in physical units reads $\sigma^2 = \hat{\sigma}_0^2 T^2 + K_n[\beta_n z/(n-1)!]^2 T^{-2(n-1)}$. Minimizing over $T$ gives $\sigma_{\min} \propto |\beta_n z|^{1/n}$, which generalizes Marcuse's $z^{1/2}$ and $z^{1/3}$ laws and agrees with the $1/B_0^N$ transmission limit of Amemiya [7]. Relative to a Gaussian pulse, the minimum achievable width is larger by the factor

$$\rho_n = \left[\frac{\hat{\sigma}_0^{2(n-1)}K_n}{\left(\hat{\sigma}_0^{2(n-1)}K_n\right)_{m=1}}\right]^{1/(2n)} \propto m^{(2n-3)/(2n)} \quad (m \gg 1). \tag{23}$$

For $n = 2$, $\rho_2 = (4\hat{\sigma}_0^2\hat{\mu}_2)^{1/4}$ involves the rms time–bandwidth product, which is bounded below by its Gaussian value, so the Gaussian is the optimum shape, in agreement with [12]. Since $\rho_n$ compares pulses at their respective optimum input widths, it does not depend on the definition of the input width. The penalty increases with $n$ (Table 2). For $m = 4$ the minimum width exceeds that of a Gaussian pulse by 26% when $\beta_2$ dominates and by 82% when $\beta_3$ dominates, and is three times larger when $\beta_6$ dominates.

**Table 2. Minimum Output Width Relative to a Gaussian Pulse, $\rho_n$[a]**

| $m$ | $\rho_2$ | $\rho_3$ | $\rho_4$ | $\rho_5$ | $\rho_6$ |
|---|---|---|---|---|---|
| 1 | 1 | 1 | 1 | 1 | 1 |
| 2 | 1.082 | 1.252 | 1.380 | 1.529 | 1.629 |
| 3 | 1.180 | 1.554 | 1.806 | 2.103 | 2.331 |
| 4 | 1.263 | 1.820 | 2.203 | 2.652 | 3.011 |

[a]Equation (23), for a narrow-band source with only $\beta_n$ present, no chirp and the input width optimized.

### *6.2 Chirp compression*

With $\beta_2$ only, Eq. (21) becomes $\sigma^2/\sigma_0^2 = 1 + \left[2DC + 4D^2\left((1+C^2)\hat{\mu}_2 + V^2/2\right)\right]/\hat{\sigma}_0^2$. The pulse is compressed when $DC < 0$, and the minimum is reached at

$$D^* = -\frac{C}{4[(1+C^2)\hat{\mu}_2 + V^2/2]}, \qquad \frac{\sigma_{\min}}{\sigma_0} = \left[1 - \frac{C^2}{4\hat{\sigma}_0^2[(1+C^2)\hat{\mu}_2 + V^2/2]}\right]^{1/2}, \tag{24}$$

with the input width recovered at $2D^*$. For $m = 1$ these expressions reduce to the results of Part 3, $\sigma_{\min}/\sigma_0 = [(1+V^2)/(1+V^2+C^2)]^{1/2}$, while the effect of the source width for $m > 1$ is new. For a monochromatic source,

$$\frac{\sigma^2}{\sigma_0^2} = 1 + \frac{2C}{\hat{\sigma}_0^2}D + \frac{4(1+C^2)\hat{\mu}_2}{\hat{\sigma}_0^2}D^2. \tag{25}$$

With $D = x/T^2$, where $x$ is the normalized distance of [12], Eq. (25) coincides with Eqs. (5) and (7) of Anderson and Lisak, $C$ playing the role of their $\alpha$, apart from the sign convention. Their results are therefore the special case $V = 0$ of the present theory with $\beta_2$ only. Since $4\hat{\sigma}_0^2\hat{\mu}_2 \geq 1$, with equality only for the Gaussian, a given chirp always compresses a super-Gaussian pulse less, and over a shorter distance. For $C = -3$ and $V = 0$ the minimum widths are 0.32, 0.59, 0.73 and 0.80 of the input width for $m$ = 1–4, reached at $D = 0.15$, 0.074, 0.049 and 0.037 (Fig. 3). The frequency sweep of steep-edged pulses is confined to the pulse edges, which contain little of the energy.

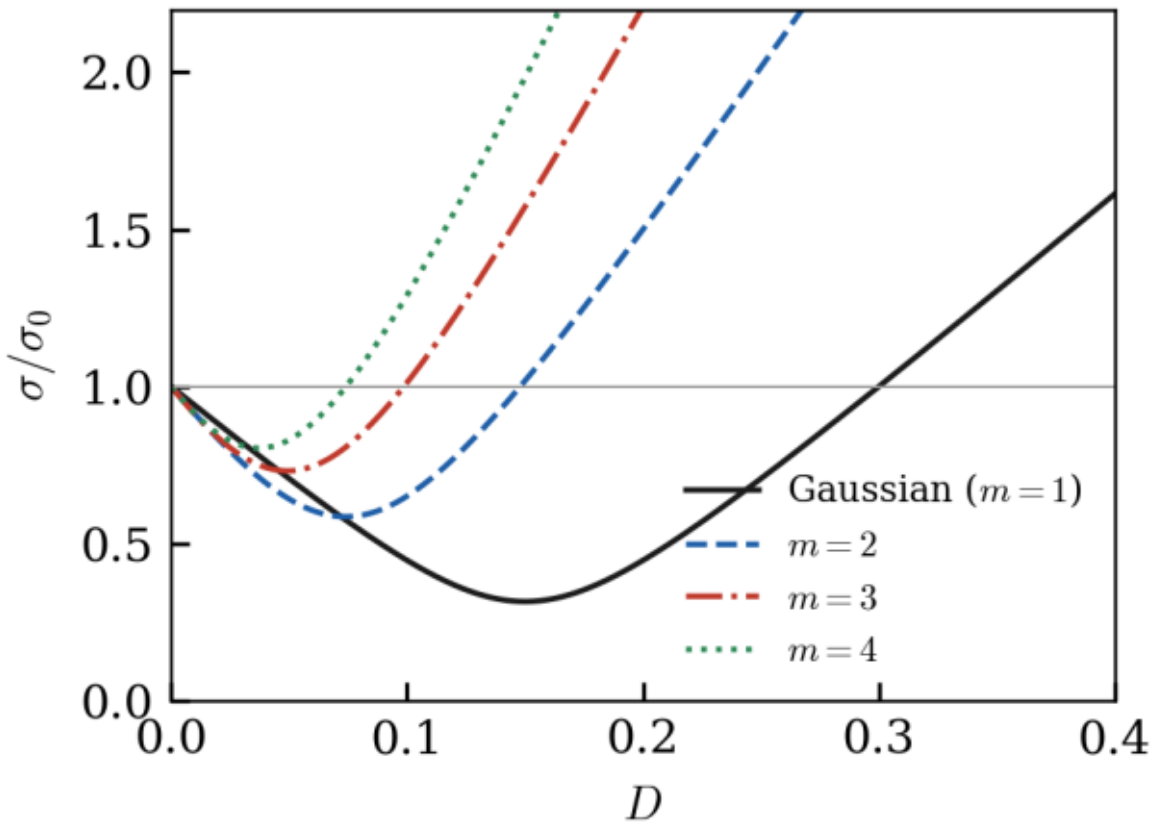

**Fig. 3.** Chirp compression with first-order dispersion only ($B = 0$, $V = 0$): rms width $\sigma/\sigma_0$ versus $D$ for $C = -3$, from Eq. (25). Line styles as in Fig. 1. The horizontal line marks the input width.

## 7. Pulse shapes and interpretation of the rms width

Figure 4 shows ensemble-averaged output pulses calculated from Eq. (3) using fast Fourier transforms, with the average over the source evaluated by Gauss–Hermite quadrature. At the zero-dispersion wavelength [Figs. 4(a) and 4(b)] the main lobe hardly depends on $m$, but the oscillatory trailing tail found by Marcuse [1] is about three times stronger for the super-Gaussian pulses; for $B < 0$ it becomes a precursor. Away from $\lambda_0$ [Figs. 4(c) and 4(d)] the super-Gaussian pulses broaden more and develop a lower, flatter top, in agreement with their larger $\hat{\mu}_2$, and a finite source width makes the output shapes more similar, as expected from remark (ii) of Section 5.3.

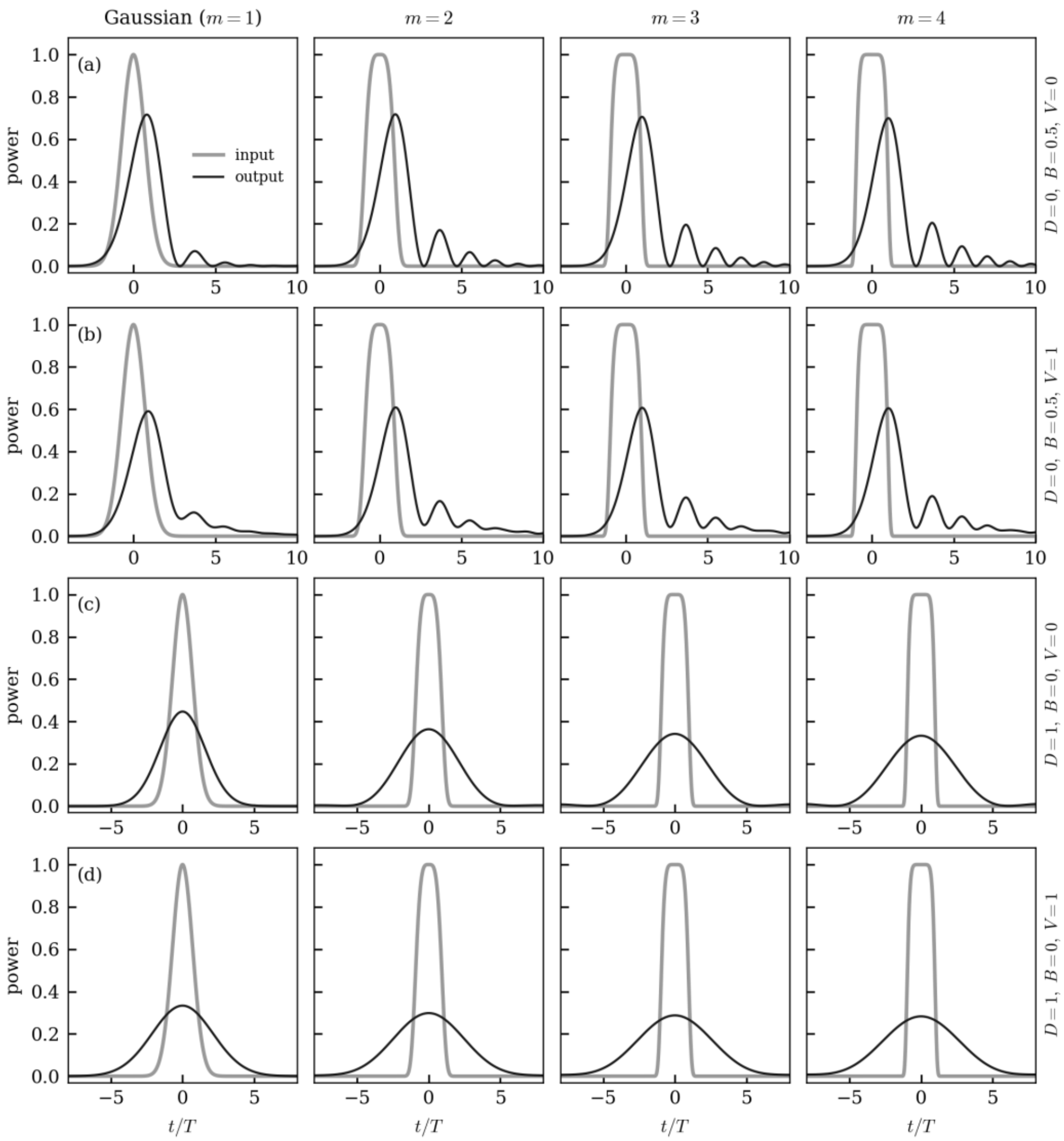

**Fig. 4.** Input and ensemble-averaged output pulses for $m = 1$ (Gaussian) to $m = 4$ (columns), computed from Eq. (3). The gray curves are the input power $|u(t)|^2$ and the black curves the output power, both normalized to the input peak and plotted in the frame moving at the group velocity. (a) Zero-dispersion wavelength ($D = 0$, $B = 0.5$) with a monochromatic source ($V = 0$). (b) As (a) with a source of width $V = 1$. (c) Away from the zero-dispersion wavelength ($D = 1$, $B = 0$), $V = 0$. (d) As (c) with $V = 1$.

Figure 5 shows the effect of dispersion beyond second order ($\beta_4$, $\beta_5$ and $\beta_6$), with each coefficient acting alone. Third- and fifth-order dispersion produce symmetric outputs with long, low skirts, whereas fourth-order dispersion produces an asymmetric pulse with an oscillatory trailing tail, in agreement with the behavior of even and odd orders described by Amemiya [7]. In all three cases the distortion increases with $m$: for $D_4 = 0.4$ the output peak falls from 0.57 of the input peak for the Gaussian pulse to 0.48 for $m = 4$, and for $D_6 = 0.1$ from 0.69 to 0.63, while the oscillatory tail produced by $D_5 = 0.2$ is markedly stronger for the super-Gaussian pulses. This is the pulse-shape counterpart of the growth of $K_n$ with $m$ discussed in Section 5.4.

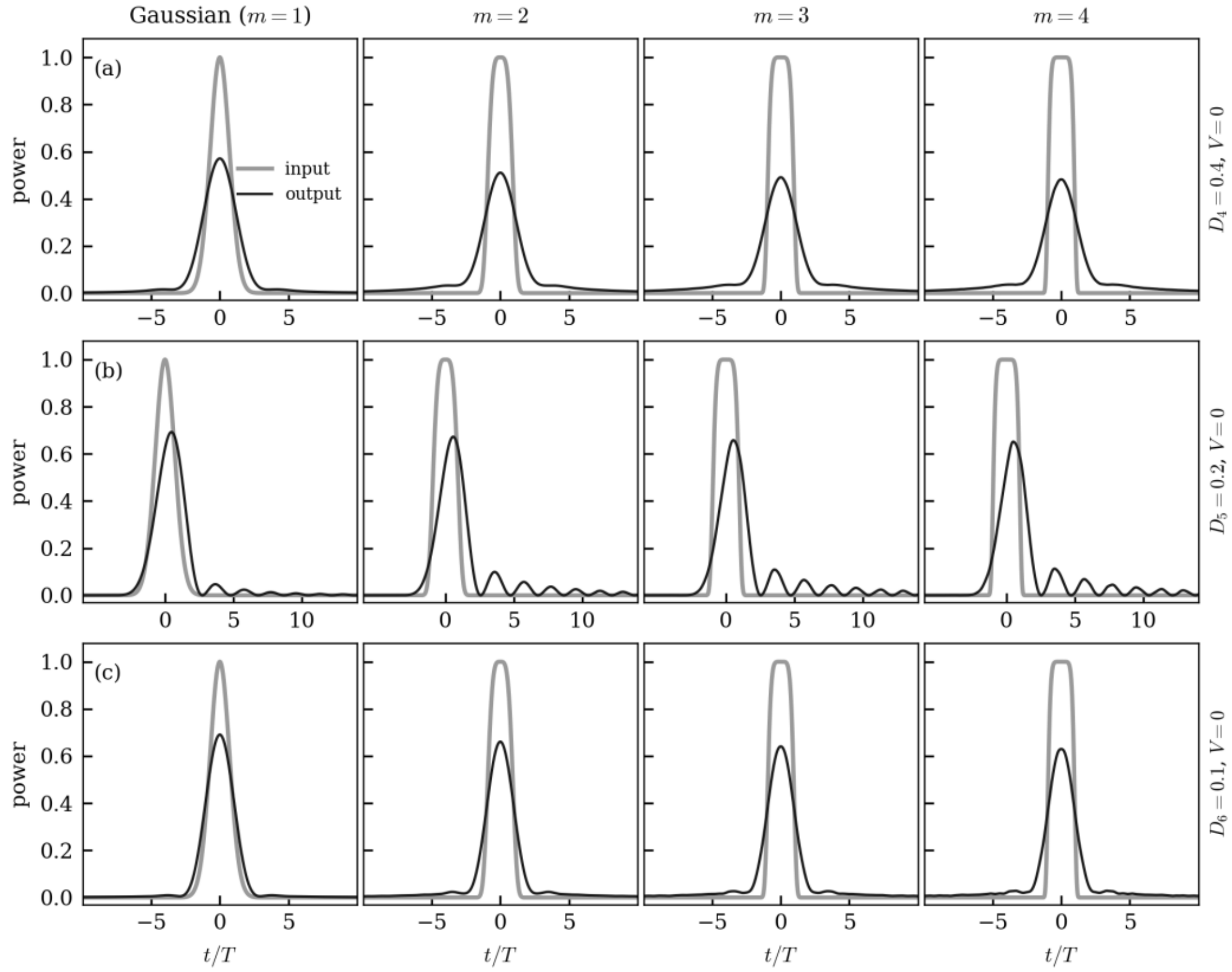

**Fig. 5.** Input and ensemble-averaged output pulses with dispersion beyond second order ($\beta_4$, $\beta_5$ and $\beta_6$), for $m = 1$ (Gaussian) to $m = 4$ (columns) and a monochromatic source ($V = 0$), computed from Eq. (3). Gray curves: input power; black curves: output power, normalized to the input peak. (a) Third-order dispersion only, $D_4 = 0.4$. (b) Fourth-order dispersion only, $D_5 = 0.2$. (c) Fifth-order dispersion only, $D_6 = 0.1$.

For these pulses the rms width must be interpreted with some care, as illustrated in Fig. 6. At the zero-dispersion wavelength and $B = 1$, the rms broadening $\sigma/\sigma_0$ is 3.16 for the Gaussian but 36.2 for $m = 4$ [Fig. 6(a)], whereas the width of the interval between the 5% and 95% points of the cumulative output energy grows by factors of only 2.80 and 9.70 [Fig. 6(b)]. The reason becomes clear from Fig. 6(c), which shows the fraction of the variance term $K_3$ of $\beta_3$ contributed by the spectral components with $|\Omega| < \Omega_c$. For the Gaussian, 99.6% comes from $|\Omega| < 3/T$, the main spectral lobe. For $m = 2$, 3 and 4 this fraction is only 28%, 10% and 5%, and frequencies up to about $10/T$–$30/T$ are needed to account for the whole term. The side lobes in Fig. 1(b) carry only a small fraction of the energy, but they travel with group delays of order $\beta_3\Omega^2 z/2$ and form a weak, distant tail that dominates the second moment. Hence, although the rms width is still a rigorous bound in the sense of Chebyshev's inequality, it is a loose one for steep-edged pulses near $\lambda_0$, and energy-containment widths or eye-opening measures are better suited for system design in this regime. Away from $\lambda_0$, where the broadening is controlled by $\hat{\mu}_2$, the rms and energy widths agree well [Table 4 and Fig. 4(c)].

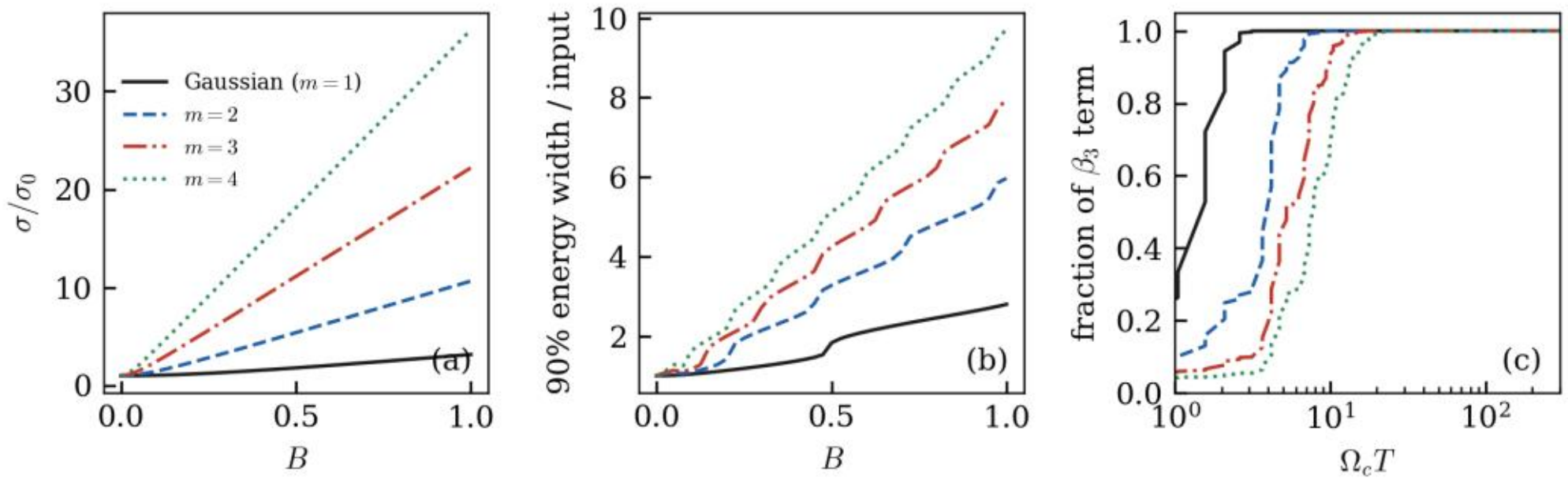

**Fig. 6.** rms width versus energy width at the zero-dispersion wavelength ($D = 0$, $V = C = 0$). (a) rms broadening $\sigma/\sigma_0$ versus $B$, from Eq. (21). (b) Width of the interval between the 5% and 95% points of the cumulative output energy, relative to its input value, computed from Eq. (3); the small steps occur when a tail oscillation crosses the 5% or 95% level. (c) Fraction of the variance term of $\beta_3$, $K_3 = \hat{\mu}_4 - \hat{\mu}_2^2$ contributed by spectral components with $|\Omega| < \Omega_c$, versus $\Omega_c T$. Line styles as in Fig. 1.

The same tail accounts for the different effects of source width and chirp at $\lambda_0$ [remark (iv) of Section 5.3]. For $m = 3$ a source width $V = 1.5$ smooths the tail, whereas an equal chirp $C = 1.5$ preserves the oscillations and, by broadening the spectral side lobes, adds a weak but far-reaching tail that raises the rms width by more than an order of magnitude without much visible change near the main lobe (Fig. 7).

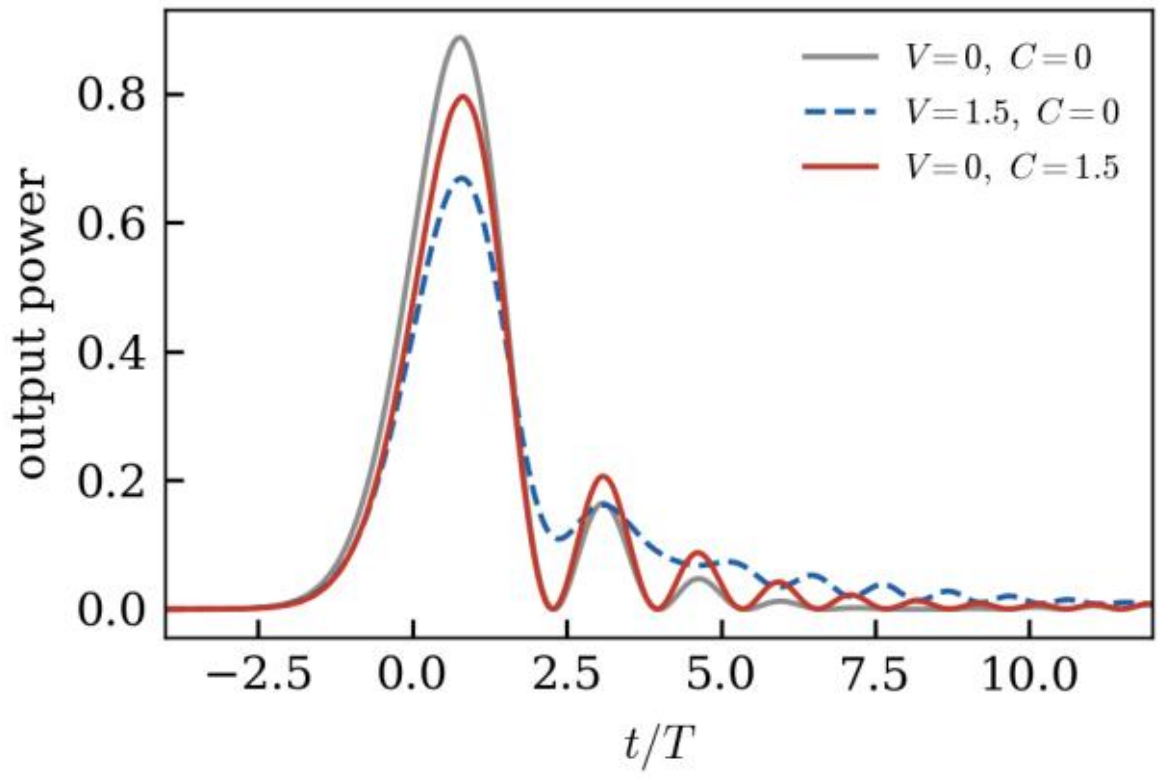

**Fig. 7.** Output pulses of a super-Gaussian of order $m = 3$ at the zero-dispersion wavelength ($D = 0$, $B = 0.3$), comparing a monochromatic unchirped input (gray), a source of width $V = 1.5$ without chirp (dashed) and a monochromatic source with chirp $C = 1.5$ (solid). For a Gaussian pulse the last two cases have the same rms width; for $m = 3$ they do not.

The distant tail also makes the numerical evaluation of the rms width demanding. Figure 8 shows the ratio of the numerically computed $\sigma^2$ to the value given by Eq. (17) as the half-width $L$ of the time window is increased. For $m = 4$ and $\beta_3$ the simulation converges at $L \approx 400T$, and with $\beta_4$ and chirp at $L \approx 3000T$. With fourth- or fifth-order dispersion less than 80% of the rms width is recovered even at $L = 3200T$, although the numerical values still approach Eq. (17). Whenever the simulation converges, it agrees with Eqs. (17)–(25) to within machine precision, and independent symbolic evaluations of the moments give the same agreement (Supplement 1).

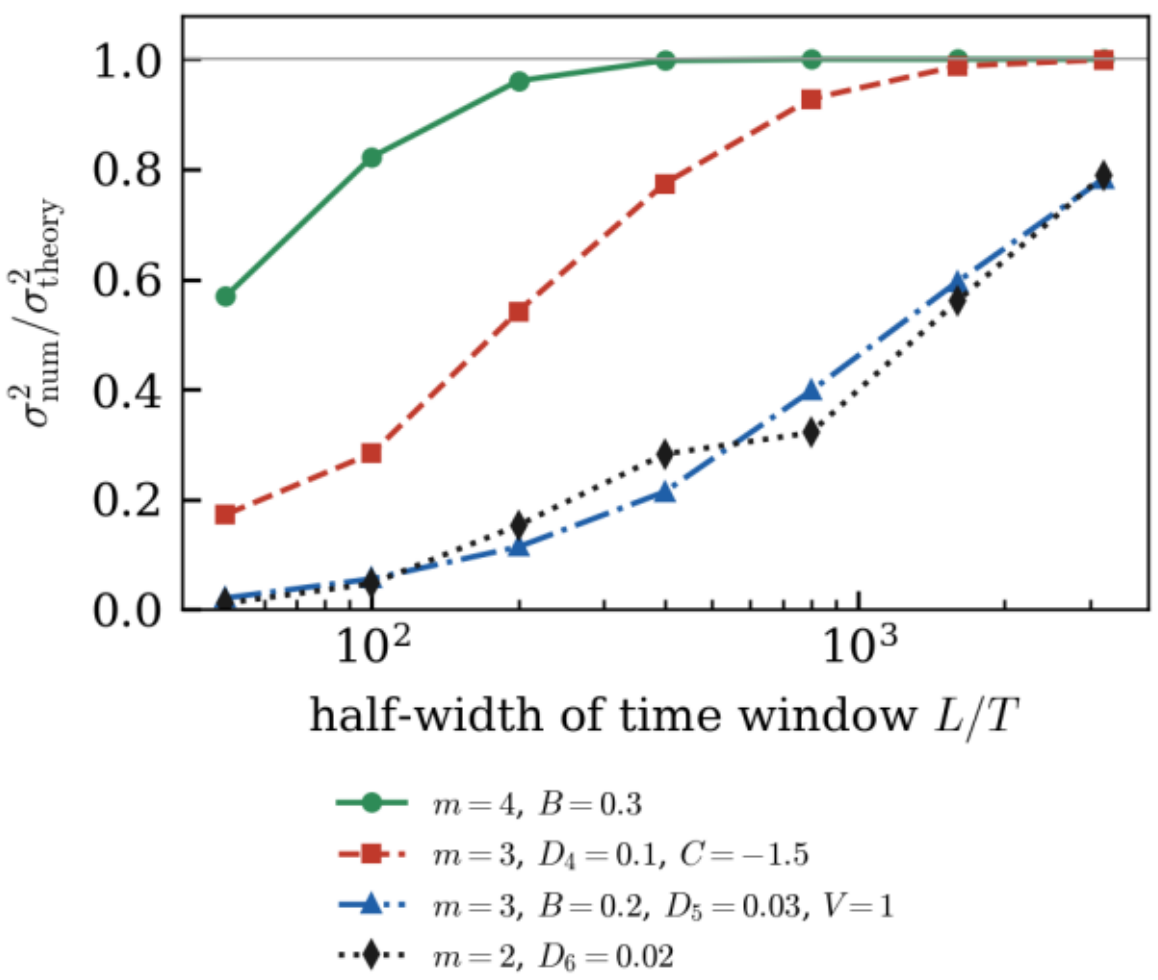

**Fig. 8.** Convergence of the numerically computed rms width with the size of the simulation window. The ratio of $\sigma^2$ computed from Eq. (3) to the exact value of Eq. (17) is plotted versus the window half-width $L/T$ for four cases combining pulse order, dispersion orders, chirp and source width (legend). The sampling interval is $0.0016\,T$ in all cases.

## 8. Pulse sequences and interference

### 8.1 Formulation

A sequence of identical pulses with bit period $\tau$ and complex amplitudes $c_k$, which describe both the bit values and the optical phases, is represented by the envelope

$$u(t) = \sum_k c_k\; u_1(t - k\tau), \tag{26}$$

where $u_1$ is the single pulse of Eq. (6). Because the modulation acts on the same source field for all pulses, each source component $\omega'$ produces the output $e^{i\omega' t}\sum_k c_k\; b(z, t - k\tau; \omega')$, with

$$b(z, t; \omega') = \frac{1}{2\pi}\int \tilde{u}_1(\nu)\, e^{-i\Delta\beta(\nu+\omega')z} e^{i\nu t} d\nu. \tag{27}$$

The carrier factor $e^{i\omega' t}$ is common to all pulses and disappears from the output power, which becomes

$$P(z,t) = \sum_k |c_k|^2\, P_1(z, t - k\tau) + 2\sum_{k<l} \mathrm{Re}\,[c_k^* c_l\, I_{l-k}(z, t - k\tau)],$$

$$I_n(z,t) = \int S(\omega')\, b^*(z,t;\omega')\, b(z, t - n\tau; \omega')\, d\omega'. \tag{28}$$

Here $P_1$ is the single-pulse output of Eq. (3), and $I_n$ is the interference term between two pulses separated by $n\tau$. The source enters $I_n$ only through $S(\omega')$, so the interference depends on the separation of the pulses and not on their absolute position in the sequence.

### 8.2 Time moments of the interference term

Propagation is described by the unitary operator $\exp[-i\Delta\beta(\hat{\Omega} + \omega')z]$, with $\hat{\Omega} = -i\, d/dt$. Since $[t, f(\hat{\Omega})] = i f'(\hat{\Omega})$, the time operator evolves as $t \to t + z\,\tau(\hat{\Omega} + \omega')$, and the time moments of the interference term are

$$\int t^p\, I_n(z,t)\, dt = \int S(\omega') \langle u_1 | [t + z\,\tau(\hat{\Omega} + \omega')]^p | u_1(\cdot - n\tau)\rangle d\omega'. \tag{29}$$

The operator in Eq. (29) contains only multiplications by $t$ and derivatives, so every term is an overlap integral between derivatives of the two input pulses. When the input pulses do not overlap, all these integrals vanish and the interference term contributes nothing to the energy, centroid or rms width of the sequence, at any distance and for any dispersion relation, chirp and source spectrum. The moments of a sequence are then those of the incoherent sum of its pulses, whatever the optical phases, and the results of Sections 3–6 apply to each pulse separately. (For $n = 0$, Eq. (29) is an alternative derivation of Eq. (11).)

The pulse shape matters here. For $\tau = 4T$ the input overlap $\int u_1(t)\,u_1(t-\tau)\,dt / \int u_1^2\,dt$ is $e^{-\tau^2/4T^2} = 0.018$ for Gaussian pulses but only $10^{-29}$ for $m = 3$. Accordingly, with $D = 2$ the energy, centroid and variance of a Gaussian pulse pair differ by up to 10% between in-phase and antiphase pulses, whereas for $m = 3$ the differences are below $10^{-15}$.

### *8.3 Gaussian pulse pair*

For two Gaussian pulses ($m = 1$, $c_1 = 1$, $c_2 = e^{i\theta}$) with first-order dispersion and a single-line source, the integrals in Eqs. (27) and (28) can be evaluated in closed form. With time in units of $T$, $\xi = \beta_2 z/T^2 = 2D$ and $\Lambda^2 = 1 + \xi^2(1 + V^2)$, the result is

$$P(z,t) = P_1(t) + P_1(t-\tau) + \frac{2}{\Lambda}\exp\left[-\frac{(t-\tau/2)^2 + (1+V^2\xi^2)\tau^2/4}{\Lambda^2}\right]$$
$$\times \cos\left[\frac{\xi\tau(t-\tau/2)}{\Lambda^2} - \theta\right],$$
$$P_1(t) = \frac{1}{\Lambda}\exp\left(-\frac{t^2}{\Lambda^2}\right). \tag{30}$$

$P_1$ is Marcuse's Gaussian solution. The interference produces fringes centered at the midpoint $t = \tau/2$, with local period $2\pi\Lambda^2/(\xi\tau)$, and their position is set by the relative phase $\theta$. Normalized to the geometric mean of the two single-pulse powers, the fringe visibility is independent of $t$,

$$\mu(z) = \exp\left[-\frac{V^2\xi^2\tau^2}{4\Lambda^2}\right] \;\rightarrow\; \mu_\infty = \exp\left[-\frac{V^2\tau^2}{4(1+V^2)}\right] \quad (\xi \rightarrow \infty). \tag{31}$$

The visibility therefore decreases from 1 at the input, where both pulses sample the same source field at each instant, to a finite limit. It does not reach the value $\exp(-V^2\tau^2/4)$ of the source coherence function at the separation $\tau$. At large distance, dispersion maps frequency onto time, so at a given output time only the source components within the pulse bandwidth contribute, and this spectral filtering partly restores the coherence. Even for a very broad source ($V \rightarrow \infty$), $\mu_\infty \rightarrow e^{-\tau^2/4}$, which is the overlap of the two input pulses.

### *8.4 Super-Gaussian pulse pairs*

The same argument applies to any pulse shape. For large $\xi$ the output at time $t$ is dominated by the frequency $t/\xi$, and the visibility at the midpoint tends to the normalized Fourier transform, at lag $\tau$, of the source spectrum filtered by the pulse power spectrum,

$$\mu_\infty = \frac{\left|\int S(\omega)|\tilde{u}_1(\omega)|^2 e^{i\omega\tau} d\omega\right|}{\int S(\omega)|\tilde{u}_1(\omega)|^2 d\omega}. \tag{32}$$

Equation (32) reduces to Eq. (31) for Gaussian pulses. For $\tau = 4T$ it gives $\mu_\infty = 0.40$, 0.058 and 0.0009 for $m = 3$ and $V = 0.5$, 1 and 2, compared with 0.45, 0.14 and 0.041 for the Gaussian, in agreement with the numerical results of Fig. 9(d). Super-Gaussian pulses from a broad source thus interfere much less than Gaussian pulses, because their input overlap is negligible.

With a narrow source the situation is reversed. Figures 9(a) and 9(b) show two pulses separated by $\tau = 4T$ after propagation with $D = 2$ and $V = 0$. The interference is strong for both shapes, and its effect on the neighboring bit slot is larger for the super-Gaussian pulses, whose broader spectrum makes them spread further into that slot. The energy that the first pulse contributes to the slot of the second pulse ($|t-\tau| < \tau/2$), including the interference term, is 0.227 of its total energy for incoherent Gaussian pulses and varies between 0.213 and 0.240 as $\theta$ changes from $\pi$ to 0, a variation of about 6%. For $m = 3$ it is 0.243 without interference and ranges from 0.180 to 0.305, about 26%. With a source of width $V = 1.5$ [Fig. 9(c)], the fringes almost disappear for $m = 3$ and the variation drops to 0.6%. Interference between neighboring pulses is therefore a phase-dependent contribution to intersymbol interference that the moment description cannot capture. It is important for steep-edged pulses from coherent sources and becomes negligible for super-Gaussian pulses from sources broader than their pulse bandwidth.

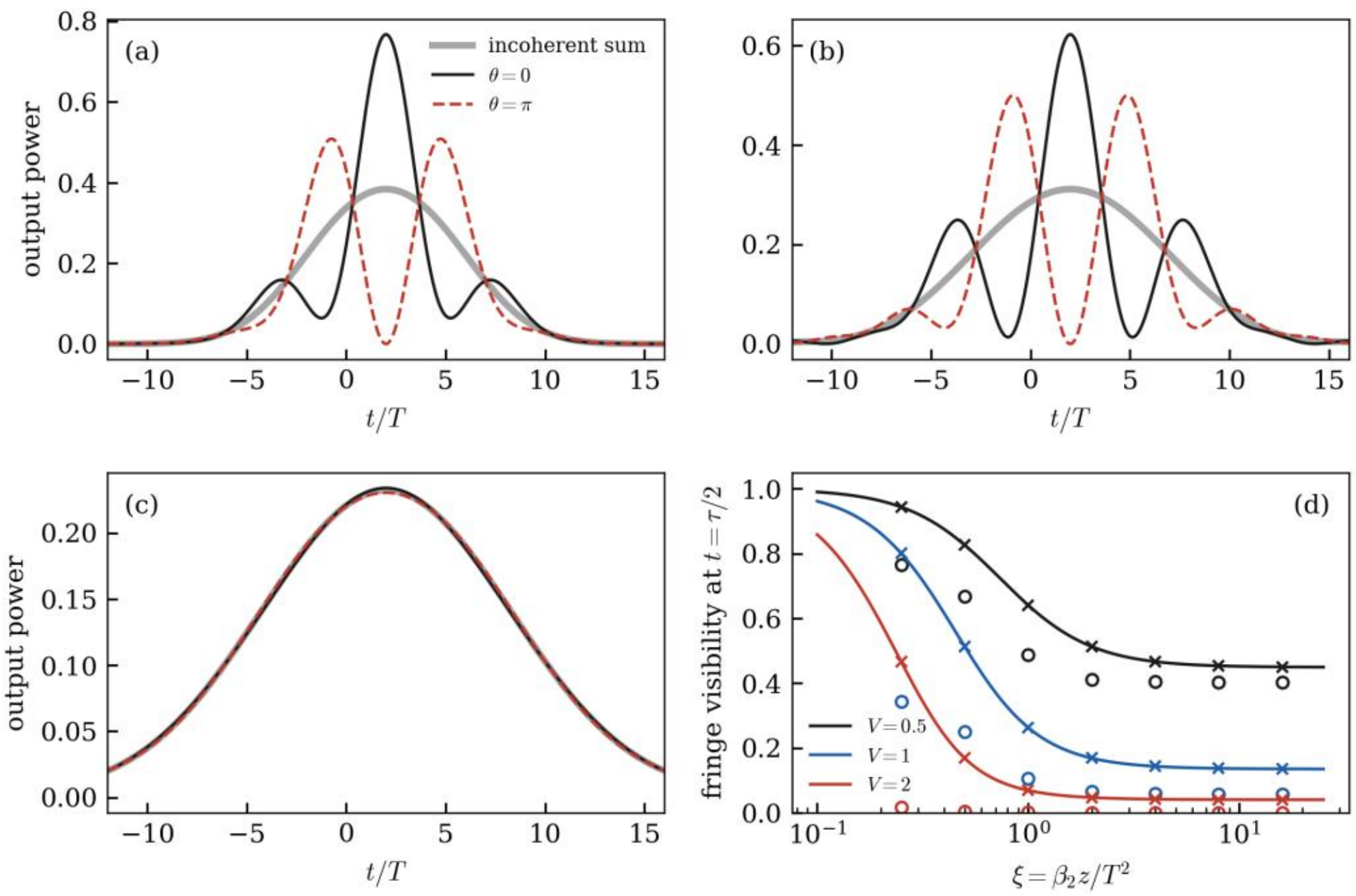

**Fig. 9.** Interference between two pulses separated by $\tau = 4T$. (a) Output power of two Gaussian pulses after propagation with $D = 2$ ($B = 0$) from a monochromatic source ($V = 0$), for in-phase ($\theta = 0$, solid) and antiphase ($\theta = \pi$, dashed) pulses; the thick gray curve is the incoherent sum. (b) The same for super-Gaussian pulses with $m = 3$. (c) As (b) with a source of width $V = 1.5$. (d) Fringe visibility at the midpoint $t = \tau/2$ versus $\xi = \beta_2 z/T^2$ for $V = 0.5$, 1 and 2: Eq. (31) for Gaussian pulses (lines), and numerical results for $m = 1$ (crosses) and $m = 3$ (circles), computed from Eq. (28).

Interference is also visible when the output pulses remain well separated. Figure 10 shows two cases with a monochromatic source in which the dip between the pulses is clearly resolved. With first-order dispersion [$D = 0.6$, $\tau = 5T$, Figs. 10(a) and 10(b)], the Gaussian pulses interfere only in the gap: the power at the midpoint, 0.15 of the peak for incoherent pulses, doubles for $\theta = 0$ and vanishes for $\theta = \pi$. The dark fringe at $t = \tau/2$ for antiphase pulses is exact for any symmetric pulse and any dispersion relation containing only coefficients $\beta_n$ with even $n$, because $b(z, t; 0)$ is then even in $t$. For $m = 3$ the input overlap is only $4 \times 10^{-108}$, yet the fringes extend over both pulses, since the spectral side lobes spread each pulse under its neighbor, and the peak power changes from 0.58 for incoherent pulses to 0.74 and 0.71 for $\theta = 0$ and $\pi$. At the zero-dispersion wavelength [$B = 0.3$, $\tau = 6T$, Figs. 10(c) and 10(d)], the oscillatory tail of the first pulse overlaps the leading edge of the second. The resulting fringes are weak for Gaussian pulses, whose peak power changes by about 3%, but for $m = 3$ they create or remove a shoulder on the leading edge of the second pulse. In all panels the energy, centroid and rms width of the pair are the same for the three curves, as required by Eq. (29), except for corrections of the order of the input overlap, which is $2 \times 10^{-3}$ and $10^{-4}$ for the Gaussian cases.

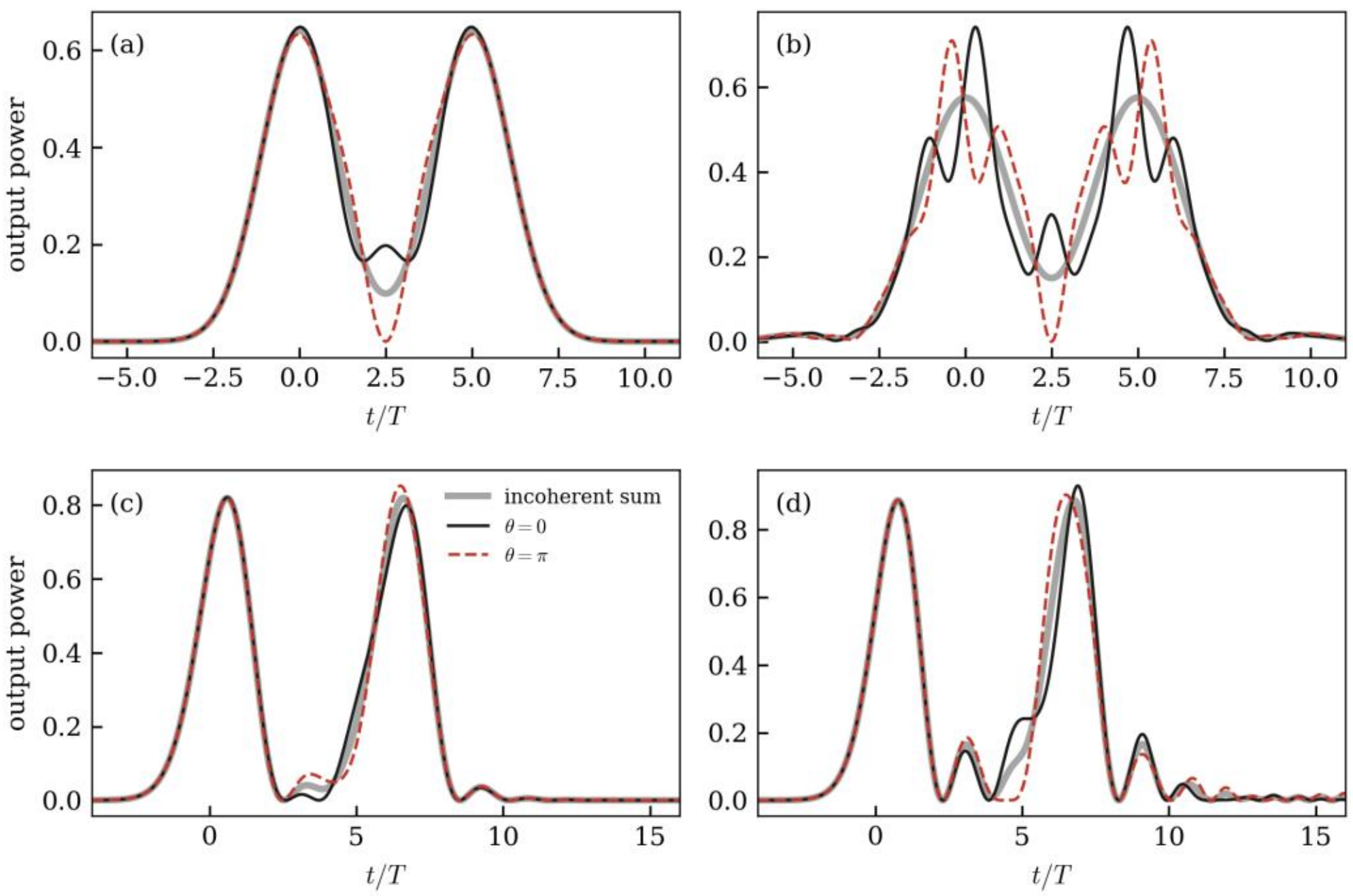

**Fig. 10.** Interference between output pulses that remain separated, for a monochromatic source ($V = 0$). (a) Gaussian pulses separated by $\tau = 5T$ after propagation with $D = 0.6$ ($B = 0$), for in-phase ($\theta = 0$, solid) and antiphase ($\theta = \pi$, dashed) pulses; the thick gray curve is the incoherent sum. (b) The same for super-Gaussian pulses with $m = 3$. (c) Gaussian pulses separated by $\tau = 6T$ at the zero-dispersion wavelength ($D = 0$, $B = 0.3$). (d) The same for $m = 3$. All curves computed from Eq. (28).

### *8.5 Eye diagrams of 100-bit sequences*

To assess the combined effect of dispersion and interference on a data signal, we computed the output for ten random on–off keyed sequences of 100 bits each, with bit period $\tau = 4T$, $c_k \in \{0,1\}$ and a common optical phase for all bits. Each sequence was treated as periodic, as produced by a pattern generator, so that the 1000 bits are free from edge effects. The output field for each source component is the superposition of the single-bit fields $b(z, t - k\tau; \omega')$ of Eq. (28), and was computed with one fast Fourier transform per sequence and source component. The traces of all bit slots, each spanning two bit periods, were superimposed to form the eye diagrams of Fig. 11. At the sampling instant that maximizes it, the eye opening EO is the difference between the lowest "one" and the highest "zero", normalized to its value at the input; the eye-opening penalty is $\mathrm{EOP} = -10\log_{10}\mathrm{EO}$. We also give the deterministic Q factor $Q = (\mu_1 - \mu_0)/(s_1 + s_0)$, where $\mu_{1,0}$ and $s_{1,0}$ are the mean and standard deviation of the ones and zeros at the sampling instant, which measures intersymbol interference in the absence of noise, and the eye width EW, the fraction of the bit period over which the eye remains open.

Table 3 summarizes the results. With first-order dispersion and a monochromatic source [Figs. 11(a) and 11(b)], consecutive ones interfere constructively at the slot boundaries, producing the bright fringes at $t = \pm\tau/2$ discussed in Section 8.4, and the eye opening is 0.35 for Gaussian and 0.23 for super-Gaussian pulses. Without interference, that is, adding the intensities of the individual bits, it would be 0.52 and 0.38, and alternating the optical phase of consecutive bits recovers 0.51 and 0.27. A source of width $V = 1.5$ suppresses the fringes but broadens the pulses further, and the opening falls to 0.21 and 0.14, now equal to the incoherent values. At the zero-dispersion wavelength [Figs. 11(e) and 11(f)] the eye is limited by the oscillatory tails, and interference reduces the opening only for $m = 3$, from 0.72 to 0.52.

Dispersion beyond second order [Figs. 11(g)–11(l)] is particularly harmful for super-Gaussian pulses. With $D_4 = 0.4$ the eye of the Gaussian pulses is still open (EO = 0.25) but that of the $m = 3$ pulses is closed, and with $D_6 = 0.1$ the openings are 0.32 and 0.01. Fourth-order dispersion, whose odd symmetry produces trailing tails similar to those of $\beta_3$, is less damaging (0.56 and 0.24). In these cases interference is the dominant impairment: without it the openings for $m = 3$ would be 0.37, 0.60 and 0.59 for $D_4$, $D_5$ and $D_6$. The deterministic Q factor and the eye width follow the same trends, and super-Gaussian pulses give a smaller eye opening than Gaussian pulses in all the cases considered.

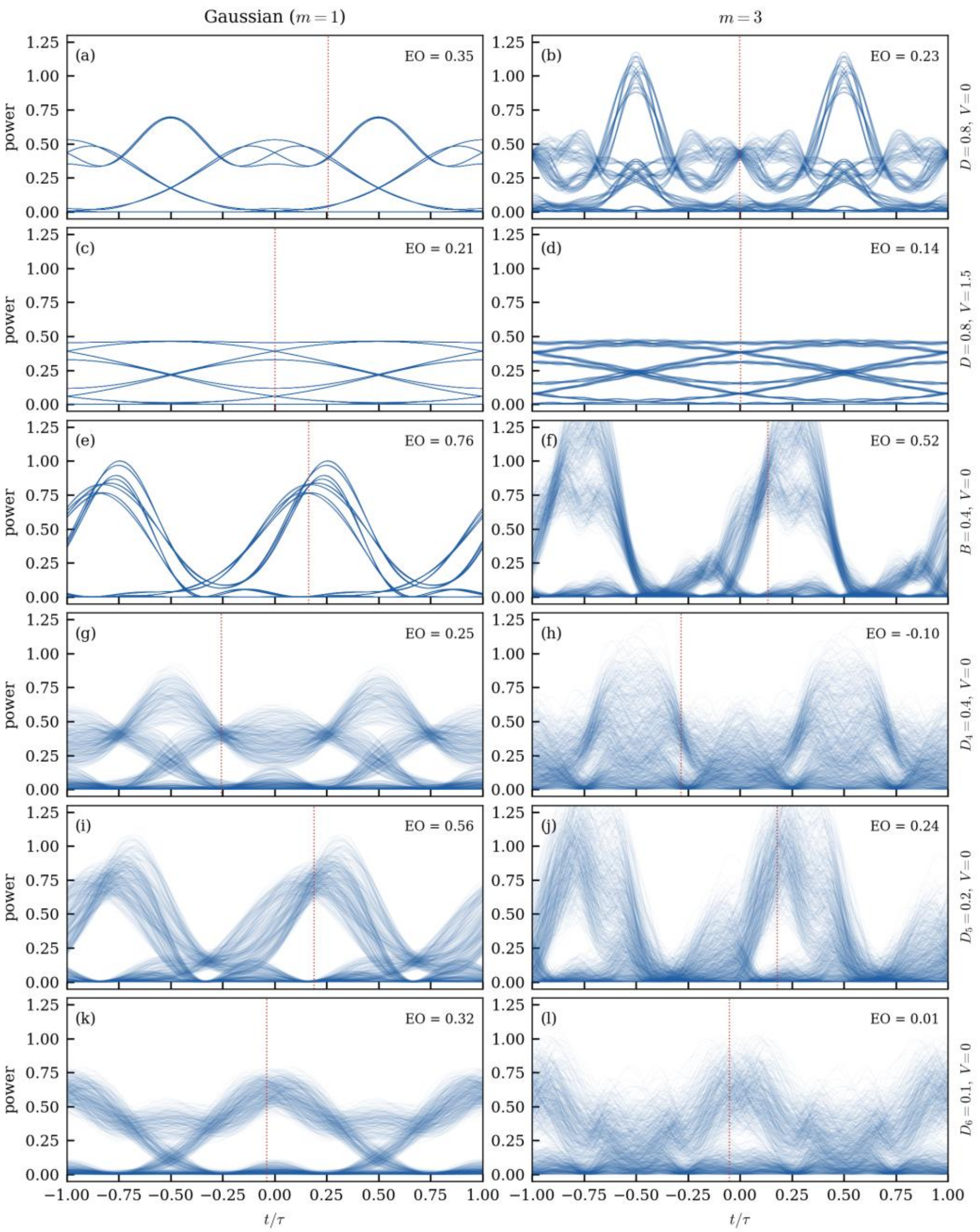

**Fig. 11.** Eye diagrams obtained from ten random periodic sequences of 100 bits (1000 bits in total) with bit period $\tau = 4T$ and all bits in phase, computed from Eq. (28). Each panel

superimposes the traces of all bit slots over two bit periods. Left column: Gaussian pulses ($m = 1$); right column: super-Gaussian pulses ($m = 3$). Rows, from top to bottom: first-order dispersion $D = 0.8$ with a monochromatic source (a, b) and with a source of width $V = 1.5$ (c, d); second-order dispersion $B = 0.4$ (e, f); third-order dispersion $D_4 = 0.4$ (g, h); fourth-order dispersion $D_5 = 0.2$ (i, j); fifth-order dispersion $D_6 = 0.1$ (k, l), all with $V = 0$ except (c, d). The dotted vertical line marks the sampling instant, and EO is the eye opening normalized to its input value.

**Table 3. Eye Metrics of 100-Bit Sequences[a]**

| Case | $m = 1$: EO | $m = 1$: EOP (dB) | $m = 1$: $Q$ | $m = 1$: EW | $m = 3$: EO | $m = 3$: EOP (dB) | $m = 3$: $Q$ | $m = 3$: EW |
|---|---|---|---|---|---|---|---|---|
| $D = 0.8, V = 0$ | 0.35 | 4.6 | 13.4 | 0.99 | 0.23 | 6.4 | 7.3 | 0.74 |
| $D = 0.8, V = 1.5$ | 0.21 | 6.8 | 4.0 | 0.98 | 0.14 | 8.5 | 3.0 | 0.82 |
| $B = 0.4, V = 0$ | 0.76 | 1.2 | 20.5 | 0.97 | 0.52 | 2.8 | 5.9 | 0.51 |
| $D_4 = 0.4, V = 0$ | 0.25 | 6.1 | 7.7 | 0.32 | closed | – | 2.6 | 0 |
| $D_5 = 0.2, V = 0$ | 0.56 | 2.5 | 9.7 | 0.47 | 0.24 | 6.2 | 3.4 | 0.19 |
| $D_6 = 0.1, V = 0$ | 0.32 | 5.0 | 5.9 | 0.83 | 0.01 | 20.6 | 2.6 | 0.02 |

[a]Ten periodic random sequences of 100 bits, $\tau = 4T$, all bits in phase. EO: eye opening normalized to its input value; EOP: eye-opening penalty; $Q$: deterministic Q factor (intersymbol interference only); EW: eye width as a fraction of the bit period.

### *8.6 Capacity*

The eye-opening penalty provides a practical capacity criterion. For each coefficient $\beta_n$ acting alone and a monochromatic source, we determined the largest normalized coefficient $D_{n,c}$ for which EOP does not exceed 1 dB, using the same 1000-bit sequences. If the pulse half-width is kept at a quarter of the bit period, $T = 1/(4B_r)$, where $B_r$ is the bit rate, Eq. (7) gives the maximum bit rate over a distance $L$,

$$B_{r,\max} = \frac{1}{4}\left[\frac{n!\,D_{n,c}}{|\beta_n|\,L}\right]^{1/n}. \tag{33}$$

The same expression applies to the rms criterion $4B_r\sigma \leq 1$, that is, $\sigma \leq T$, with $D_{n,c}$ replaced by the value at which the rms width of Eq. (17) reaches $T$, $D_{n,c}^{\text{rms}} = [(1 - \hat{\sigma}_0^2)/(n^2 K_n)]^{1/2}$. Figure 12(c) compares both critical coefficients for $n = 2$–6 and $m = 1$–4. For first-order dispersion the two criteria agree to within a factor of about 1.4, and super-Gaussian pulses tolerate slightly more dispersion than Gaussian pulses under the eye criterion ($D_{2,c} = 0.41$ for $m = 2$ and 0.39 for $m = 3$, against 0.36 for the Gaussian), because their flat tops keep more energy in the bit slot. For higher orders the rms criterion becomes increasingly pessimistic for super-Gaussian pulses: for $n = 6$ and $m = 4$ it underestimates the tolerable dispersion by a factor of about 585, a direct consequence of the remote spectral tails discussed in Section 7. Figures 12(a) and 12(b) show the resulting capacity curves for standard single-mode fiber. In the C band both pulse shapes allow about 15 Gb/s over 10 km under the eye criterion, whereas at the zero-dispersion wavelength the eye criterion gives about 330 Gb/s for Gaussian and 310 Gb/s for $m = 3$ pulses over 10 km, and the rms criterion would reduce the latter to 200 Gb/s. The $L^{-1/n}$ dependence of Eq. (33) is the scaling law of Section 6.1 and of [7].

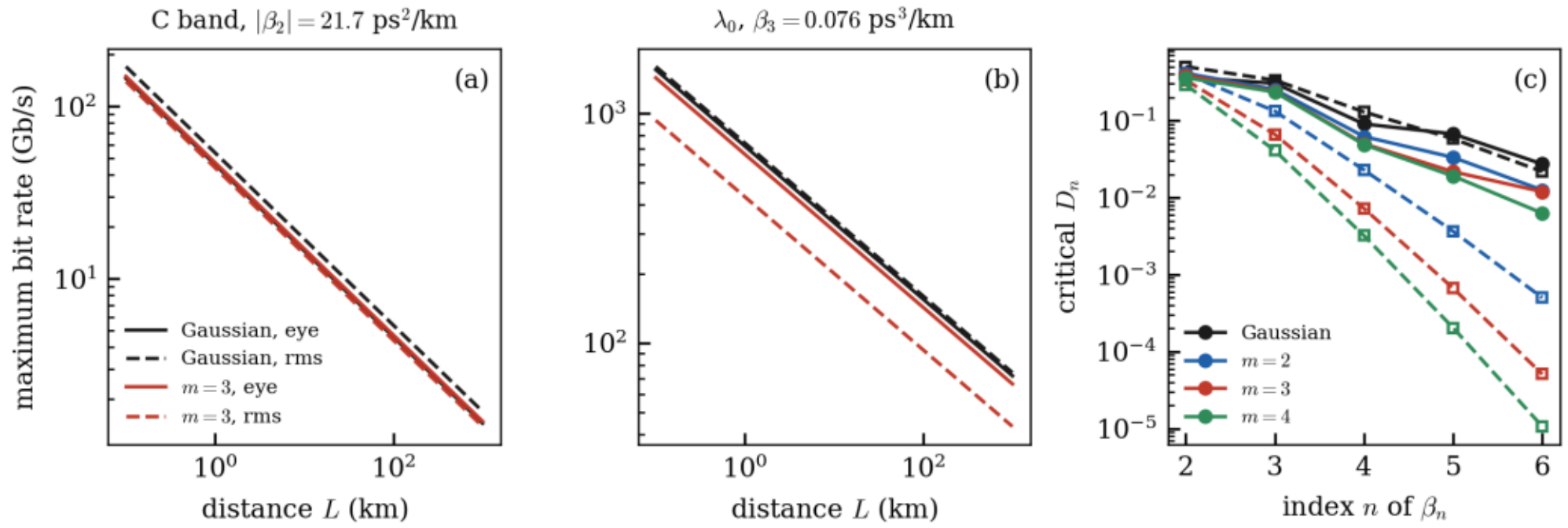

**Fig. 12.** Dispersion-limited capacity with RZ pulses of half-width $T = 1/(4B_r)$ and a monochromatic source. (a) Maximum bit rate versus distance in the C band of a standard single-mode fiber ($|\beta_2| = 21.7$ ps²/km) for Gaussian (black) and $m = 3$ (red) pulses, from Eq. (33) with the 1-dB eye-opening-penalty criterion (solid) and the rms criterion $\sigma \le T$ (dashed). (b) The same at the zero-dispersion wavelength ($\beta_3 = 0.076$ ps³/km). (c) Critical normalized coefficient $D_{n,c}$ versus the index $n$ of the dispersion coefficient $\beta_n$ for $m = 1$–4, from the eye criterion (filled circles, solid lines) and the rms criterion (open squares, dashed lines).

## 9. Examples for standard single-mode fiber

As an illustration, Table 4 gives results for a standard single-mode fiber compliant with ITU-T G.652 [16]. In the C band we take $D = 17$ ps/(nm·km), so $\beta_2 = -21.7$ ps²/km at 1550 nm. At the zero-dispersion wavelength we take the maximum dispersion slope $S_0 = 0.092$ ps/(nm²·km) at 1310 nm, which gives $\beta_3 = \lambda_0^4 S_0/(2\pi c)^2 = 0.076$ ps³/km. A narrow-linewidth laser is assumed, so that $V$ is negligible, and Gaussian and third-order super-Gaussian pulses with the same $1/e$ half-width $T$ are compared.

**Table 4. Pulse Broadening in a Standard Single-Mode Fiber[a]**

| Operating point | $T$ (ps) | $z$ (km) | $D$ | $B$ | $m = 1$: $\sigma/\sigma_0$ | $m = 1$: $E_{90}$ | $m = 3$: $\sigma/\sigma_0$ | $m = 3$: $E_{90}$ |
|---|---|---|---|---|---|---|---|---|
| C band, 1550 nm | 5 | 2 | -0.867 | 0 | 2.002 | 2.002 | 3.921 | 3.467 |
| C band, 1550 nm | 1 | 0.1 | -1.084 | 0 | 2.388 | 2.388 | 4.843 | 4.290 |
| $\lambda_0$ = 1310 nm | 5 | 40 | 0 | 0.0041 | 1.000 | 1.000 | 1.004 | 0.995 |
| $\lambda_0$ = 1310 nm | 2 | 20 | 0 | 0.0318 | 1.005 | 1.004 | 1.224 | 1.122 |
| $\lambda_0$ = 1310 nm | 0.5 | 1 | 0 | 0.1018 | 1.046 | 1.042 | 2.469 | 1.171 |

[a]ITU-T G.652 fiber [16]; $D$ and $B$ from Eq. (7). $\sigma/\sigma_0$ from Eq. (21); $E_{90}$ is the ratio of the output to input widths containing 90% of the energy (5%–95% points), computed from Eq. (3).

In the C band the broadening is dominated by $\beta_2$, and the rms and energy widths agree to within about 13% for both shapes. The super-Gaussian pulse broadens roughly twice as much as a Gaussian pulse of equal $1/e$ width, as expected from its larger $\hat{\mu}_2$. At the zero-dispersion wavelength, second-order dispersion is negligible for 5-ps pulses over 40 km but becomes important for picosecond pulses. For 0.5-ps flat-top pulses over 1 km the Gaussian pulse broadens by about 4%, whereas the rms width of the $m = 3$ pulse increases by a factor of 2.5 and its energy width by 17%. The large difference between these two figures is a direct consequence of the spectral tail discussed in Section 7.

## 10. Relation to previous work

The Gaussian case with $\beta_3$ and source bandwidth was treated by Miyagi and Nishida [4,5] and by Marcuse [1,3], and extended to arbitrary dispersion in [6]; Section 5.2 gives the corresponding rms width in explicit closed form. Amemiya derived transmission limits for each dispersion coefficient [7], consistent with the $|\beta_n z|^{1/n}$ scaling of Section 6.1, and Helczynski et al. studied the effect of higher-order dispersion on the broadening of partially incoherent light [14]. In the present work the pulse shape is added as a further variable in all these problems.

The work most closely related to ours is that of Anderson and Lisak [12]. They showed that the rms width of a coherent pulse varies parabolically with distance for any pulse form and chirp, pointed out that this remains true at higher dispersion orders, and evaluated the coefficients for $\beta_2$ and the edge-chirped super-Gaussian pulse of Agrawal and Potasek [10]. From these coefficients they obtained the bit-rate–distance product and the linear compression of super-Gaussian pulses, and they also considered the linearly chirped, nearly rectangular pulses of two-stage compressors. Some of our results coincide with theirs, namely the parabolic dependence on distance, the Gamma-function factors of the $\beta_2$ coefficients, the optimality of the Gaussian shape under $\beta_2$, and Eq. (25), which reproduces their coefficients. The main differences are summarized in Table 5 and discussed below.

First, the source width $V$ and multi-line spectra are included on the same footing as pulse shape and chirp. This leads to the source terms in Eqs. (18) and (20), the source contribution to the compression law of Eq. (24), the shape-independent limit for broad sources, and the coupling between dispersion orders produced by asymmetric line sets.

Second, although Anderson and Lisak noted that the parabolic form holds at higher orders, they gave the coefficients only for $\beta_2$. Here both coefficients are obtained for any number of dispersion terms, with chirp and a finite source width, as finite sums of Gamma functions [Eqs. (13), (17) and (18)], together with the selection rules and the asymptotic behavior for large $m$.

Third, higher-order dispersion considerably strengthens the conclusion of [12] that the Gaussian shape is superior. The $\beta_3$ broadening of super-Gaussian pulses is larger by factors of 12.5–146 for $m = 2$–4, and the optimum-width penalty $\rho_n$ reaches 82% for $\beta_3$ and 201% for $\beta_6$ at $m = 4$. Since $\rho_n$ compares pulses at their respective optimum widths, this comparison is free from the dependence on the width definition that Anderson and Lisak regarded as making it "somewhat arbitrary."

Fourth, we have obtained the coupling between chirp and second-order dispersion and have shown that Marcuse's equivalence of source width and chirp at $\lambda_0$ does not hold for non-Gaussian pulses, that third-order dispersion contributes to the linear coefficient, and that chirp compresses steep-edged pulses less efficiently.

Finally, the ensemble-averaged pulse shapes and the analysis of the spectral origin of the rms width show that near $\lambda_0$ the rms width overestimates the broadening of the energy-carrying part of the pulse, by a factor of about 4 for $m = 4$ and $B = 1$. This does not affect the $\beta_2$ results of [12], for which the rms width remains representative. In addition, the extension to pulse sequences in Section 8 shows that interference between pulses is invisible to the moment description when the input pulses do not overlap, and quantifies it through the fringe visibility of Eqs. (31) and (32).

**Table 5. Scope of Ref. [12] Compared with the Present Work**

| | Anderson and Lisak [12] | This work |
|---|---|---|
| Pulse | Super-Gaussian with edge chirp [10]; linear chirp for $m \gg 1$ | Super-Gaussian with the same chirp for all $m$ (Marcuse's linear chirp for $m = 1$); large-$m$ asymptotics |
| Source | Coherent (monochromatic) | Single- or multi-line Gaussian spectrum of any width |
| Dispersion | $\beta_2$ (parabolic law noted to hold at any order) | Any number of coefficients, explicit Gamma-function coefficients |
| Transmission figure | Bit-rate–distance product at fixed $1/e$ width | Minimum width at optimized input width, $\rho_2$–$\rho_6$ |
| Chirp | Compression under $\beta_2$ | Plus source width, chirp–$\beta_3$ coupling, $\beta_4$ compression |
| Gaussian limit | $m = 1$, $\beta_2$ | Explicit closed form for any dispersion order; exact reduction to Marcuse |

| | Anderson and Lisak [12] | This work |
|---|---|---|
| Pulse shapes | Not computed (compared with numerics of [10]) | Ensemble-averaged shapes; spectral origin of rms width; energy widths |
| Pulse sequences | Not considered | Interference term, invariance of moments, fringe visibility |

## 11. Conclusions

We have extended Marcuse's theory of pulse distortion in single-mode fibers to chirped super-Gaussian input pulses of arbitrary order, single- or multi-line Gaussian sources and an arbitrary number of dispersion coefficients. The rms output width follows exactly from the moment relation of Eq. (11) for any dispersion relation and source spectrum. The moments required are finite sums of Gamma functions, and their behavior for large $m$ is described by the tangent numbers. The results reduce to those of Marcuse for Gaussian pulses and to those of Anderson and Lisak for first-order dispersion, and they are in agreement with independent symbolic and numerical calculations.

For narrow-band sources, super-Gaussian pulses are much more sensitive than Gaussian pulses to higher-order dispersion. Their minimum achievable width exceeds the Gaussian value by 8–26% under $\beta_2$, 25–82% under $\beta_3$ and 63–201% under $\beta_6$ for $m$ = 2–4, whereas for broad sources the broadening becomes independent of pulse shape. Chirp compression, by $\beta_2$ or $\beta_4$, is weaker, and Marcuse's equivalence of source width and chirp at $\lambda_0$ holds only for Gaussian pulses. Near $\lambda_0$, and more so for higher dispersion orders, the rms width of steep-edged pulses is determined by weak spectral side lobes; it is still a rigorous bound, but it overestimates the broadening of the pulse energy. In standard fiber this is relevant for picosecond flat-top pulses near the zero-dispersion wavelength, but not in the C band, where first-order dispersion dominates.

The formalism extends directly to pulse sequences. When the input pulses do not overlap, which is very nearly the case for super-Gaussian pulses, interference between neighboring pulses does not change any time moment of the sequence, but it redistributes energy locally. The fringe visibility falls from unity to a limit given by the Fourier transform of the source spectrum filtered by the pulse spectrum, so that dispersion partly restores the coherence between pulses. For coherent sources the resulting phase-dependent intersymbol interference is stronger for super-Gaussian than for Gaussian pulses, whereas for broad sources it is much weaker. Eye diagrams of 1000-bit sequences show that interference between consecutive ones reduces the eye opening, that dispersion beyond second order can close the eye of super-Gaussian pulses while that of Gaussian pulses remains open, and that super-Gaussian pulses give smaller eye openings in all the cases considered. Capacity curves based on a 1-dB eye-opening penalty show that the rms criterion is adequate for first-order dispersion but increasingly pessimistic for super-Gaussian pulses at higher orders.

## Appendix A: Incoherent superposition

Let the source field be $E_s(t) = e^{i\omega_0 t}\int \tilde{e}(\omega')e^{i\omega' t}d\omega'$, a stationary random process with $\langle \tilde{e}(\omega_1)\tilde{e}^*(\omega_2)\rangle = S(\omega_1)\delta(\omega_1-\omega_2)$. The modulator multiplies it by the deterministic envelope $u(t)$, independent of the source fluctuations. Each source component then produces the output $\tilde{e}(\omega')a(z,t;\omega')$, with $a$ as in Eq. (3). The output power averaged over the ensemble is

$$\langle |E(z,t)|^2\rangle = \iint \langle \tilde{e}(\omega_1)\tilde{e}^*(\omega_2)\rangle a(z,t;\omega_1)a^*(z,t;\omega_2)\, d\omega_1 d\omega_2 = \int S(\omega')|a(z,t;\omega')|^2 d\omega', \tag{A1}$$

which is Eq. (3). The derivation assumes that the source is stationary, so that different spectral components are uncorrelated, and that the modulation is independent of the source fluctuations, as in Marcuse's treatment [1]. The corresponding mutual coherence function has the Schell form $u(t_1)u^*(t_2)\gamma(t_2-t_1)$, where $\gamma$ is the Fourier transform of $S$, a special case of the spectrally partially coherent pulses of [8].

## Appendix B: Weyl-ordered moments

Multiplying Eq. (10) by $\Omega^l$ and integrating over $\Omega$ gives $\int \Omega^l \mathcal{W}\, d\Omega = (-i\,\partial_s)^l [u(t+s/2)u^*(t-s/2)]_{s=0}$, and the Leibniz rule yields

$$\int \Omega^l \mathcal{W}(t,\Omega)\, d\Omega = \left(\frac{-i}{2}\right)^l \sum_{k=0}^{l} \binom{l}{k} (-1)^{l-k} u^{(k)}(t)\, u^{*(l-k)}(t). \quad \text{(B1)}$$

With $u^{(k)} = P_k u$ from Eq. (12), multiplying by $t^j$ and integrating over $t$ gives Eq. (13), since $|u|^2 = e^{-\hat{t}^{2m}}$ for every chirp. The result is real because the sum is Hermitian under $k \to l-k$. For the unchirped pulse and $j = l = 2$, Eq. (B1) gives $\langle \hat{t}^2 \hat{\Omega}^2 \rangle_W = \mathcal{I}[\hat{t}^2 P_1^2] - 1/2$, and with $P_1 = -m\hat{t}^{2m-1}$ and $\mathcal{I}[\hat{t}^{4m}] = \Gamma(2+a)/\Gamma(a)$ this equals $m^2 a(1+a) - 1/2 = (2m-1)/4$.

## Appendix C: Large-order asymptotics

Near the edge $\hat{t} = 1$, write $\hat{t} = 1 + s/(2m)$. For large $m$, $\hat{t}^{2m} \to e^s$, so $u \to f(s) = \exp(-e^s/2)$ and $d/d\hat{t} = 2m\, d/ds$. Each edge then contributes $(2m)^{2k-1} \int \left|f^{(k)}(s)\right|^2 ds$ to $\int \left|u^{(k)}\right|^2 d\hat{t}$, while $\int |u|^2 d\hat{t} \to 2$. Hence

$$\hat{\mu}_{2k} \simeq (2m)^{2k-1} \int_{-\infty}^{\infty} \left|f^{(k)}(s)\right|^2 ds = \frac{\mathcal{T}_k}{2}\, m^{2k-1}. \quad \text{(C1)}$$

With $y = e^s$, $f^{(k)}$ becomes a polynomial in $y$ multiplied by $e^{-y/2}$, and the integral reduces to Gamma functions of integer argument. Exact evaluation gives $\mathcal{T}_k/2$ for $k = 1$–$5$, where $\mathcal{T}_k = 1$, 2, 16, 272 and 7936 are the tangent numbers, and we conjecture that the same holds for all $k$.

**Funding.** [To be completed by the author.]

**Disclosures.** The author declares no conflicts of interest.

**Data availability.** Data underlying the results presented in this paper are not publicly available at this time but may be obtained from the author upon reasonable request. The code used to generate all figures and tables is available from the author.

**Supplemental document.** See Supplement 1 for supporting content.

# Supergaussian pulse distortion in singlemode fibers arbitrary dispersion: supplementary material

This supplement contains the numerical checks of the analytical results of the main paper, whose equation numbers are used throughout. The numerical results were obtained by evaluating the ensemble average of Eq. (3) with fast Fourier transforms on a time window of half-width $L$ sampled with $N$ points, and averaging over the source spectrum by Gauss–Hermite quadrature. For polynomial dispersion of order $N_d$, the second moment $\langle t^2 \rangle_{\omega'}$ is a polynomial of degree $2(N_d - 1)$ in the source frequency $\omega'$, and the quadrature gives the rms width exactly with $N_d$ or more nodes.

## 1. First- and second-order dispersion

In Table S1, Eq. (21) is compared with the numerically computed rms width for $m$ = 1–4, first- and second-order dispersion of both signs, source widths up to $V = 2$ and chirps up to $|C| = 2$. With $L = 3000T$ and $N = 2^{22}$ the relative discrepancy is below $10^{-11}$ in all unchirped cases and in all chirped Gaussian cases. For chirped super-Gaussian pulses the chirp broadens the spectral side lobes, the delayed tail extends further, and the remaining discrepancy (up to $9.4 \times 10^{-3}$ for $m = 4$, $C = -2$) is due to the window only: for that case it falls to $3.4 \times 10^{-5}$ with $L = 8000T$ and to $4 \times 10^{-8}$ with $L = 16000T$. Table S4 illustrates how a smaller window leads to an underestimate of the rms width for a steep-edged pulse at the zero-dispersion wavelength.

**Table S1. rms Width with First- and Second-Order Dispersion: Eq. (21) Versus Numerical Evaluation of Eq. (3)**

| $m$ | $D$ | $B$ | $V$ | $C$ | Eq. (21) | Numerical | Relative error |
|---|---|---|---|---|---|---|---|
| 1 | 0.5 | 0 | 0 | 0 | 2.000000 | 2.000000 | $4.4 \times 10^{-16}$ |
| 1 | 0 | 0.3 | 0 | 0 | 1.810000 | 1.810000 | $2.5 \times 10^{-16}$ |
| 1 | 0.4 | 0.2 | 1.5 | 0 | 6.882500 | 6.882500 | $9.0 \times 10^{-16}$ |
| 1 | -0.3 | 0.25 | 0.8 | 1.2 | 6.004900 | 6.004900 | $7.4 \times 10^{-16}$ |
| 1 | 0.2 | -0.3 | 0 | -2 | 20.450000 | 20.450000 | $6.9 \times 10^{-16}$ |
| 1 | 0 | 0.2 | 2 | 1 | 13.960000 | 13.960000 | $1.1 \times 10^{-15}$ |
| 2 | 0.5 | 0 | 0 | 0 | 4.000000 | 4.000000 | $< 10^{-16}$ |
| 2 | 0 | 0.3 | 0 | 0 | 11.117825 | 11.117825 | $6.4 \times 10^{-16}$ |
| 2 | 0.4 | 0.2 | 1.5 | 0 | 17.103150 | 17.103150 | $1.0 \times 10^{-15}$ |
| 2 | -0.3 | 0.25 | 0.8 | 1.2 | 84.373772 | 84.373772 | $6.7 \times 10^{-16}$ |
| 2 | 0.2 | -0.3 | 0 | -2 | 541.561913 | 541.561913 | $4.9 \times 10^{-12}$ |
| 2 | 0 | 0.2 | 2 | 1 | 57.569706 | 57.569706 | $1.2 \times 10^{-16}$ |
| 3 | 0.5 | 0 | 0 | 0 | 5.776374 | 5.776374 | $3.1 \times 10^{-16}$ |
| 3 | 0 | 0.3 | 0 | 0 | 45.234532 | 45.234532 | $3.1 \times 10^{-16}$ |
| 3 | 0.4 | 0.2 | 1.5 | 0 | 36.577269 | 36.577269 | $1.2 \times 10^{-15}$ |
| 3 | -0.3 | 0.25 | 0.8 | 1.2 | 372.810347 | 372.810345 | $5.4 \times 10^{-9}$ |
| 3 | 0.2 | -0.3 | 0 | -2 | 2566.172027 | 2565.886378 | $1.1 \times 10^{-4}$ |
| 3 | 0 | 0.2 | 2 | 1 | 180.995600 | 180.995600 | $3.6 \times 10^{-12}$ |
| 4 | 0.5 | 0 | 0 | 0 | 7.435581 | 7.435581 | $< 10^{-16}$ |
| 4 | 0 | 0.3 | 0 | 0 | 119.119248 | 119.119248 | $3.0 \times 10^{-12}$ |
| 4 | 0.4 | 0.2 | 1.5 | 0 | 73.226447 | 73.226447 | $1.6 \times 10^{-15}$ |
| 4 | -0.3 | 0.25 | 0.8 | 1.2 | 1007.105468 | 1007.070041 | $3.5 \times 10^{-5}$ |

| $m$ | $D$ | $B$ | $V$ | $C$ | Eq. (21) | Numerical | Relative error |
|---|---|---|---|---|---|---|---|
| 4 | 0.2 | -0.3 | 0 | -2 | 7074.169407 | 7008.006936 | $9.4 \times 10^{-3}$ |
| 4 | 0 | 0.2 | 2 | 1 | 440.215865 | 440.215627 | $5.4 \times 10^{-7}$ |

## 2. Higher-order dispersion

The moments $\langle \hat{X}^n \rangle$ and $\langle \hat{t}\hat{X}^n \rangle$ that enter Eq. (17) are also equal to time-domain integrals of symbolic derivatives of the chirped pulse, $\mathrm{Re}\int u^*(-i\,d/dt)^n u\,dt$ and $\mathrm{Re}\int u^*(-i\,d/dt)^n (tu)\,dt$. We evaluated these by 30-digit quadrature for $m = 1$–4 and $C = 0$, 1.5 and $-2$, up to $n = 10$, which is sufficient for dispersion up to fifth order. They agree with Eqs. (13) and (18) to at least 23 significant digits. Table S2 compares Eq. (17) with the direct numerical evaluation of Eq. (3) for those cases in which the simulation converges within a window of $L = 3000T$ to better than $10^{-4}$. Cases that do not converge are discussed in Section 7 of the main paper (Fig. 8).

**Table S2. rms Width with Dispersion up to Fifth Order: Eq. (17) Versus Numerical Evaluation of Eq. (3)**

| $m$ | Coefficients | $V$ | $C$ | Eq. (17) | Numerical | Relative error |
|---|---|---|---|---|---|---|
| 1 | $D_2 = 0.3$, $D_4 = 0.05$ | 0.5 | 1 | 7.89109 | 7.89109 | $2.3 \times 10^{-16}$ |
| 1 | $D_4 = 0.1$ | 0 | -1.5 | 15.7469 | 15.7469 | $6.8 \times 10^{-16}$ |
| 1 | $D_3 = 0.2$, $D_5 = 0.03$ | 1 | 0 | 11.08 | 11.08 | $4.8 \times 10^{-16}$ |
| 1 | $D_2 = 0.2$, $D_3 = 0.1$, $D_4 = 0.05$, $D_5 = 0.02$, $D_6 = 0.01$ | 0.7 | 0.5 | 13.946 | 13.946 | $7.0 \times 10^{-14}$ |
| 1 | $D_6 = 0.02$ | 0 | 0 | 1.8505 | 1.8505 | $9.6 \times 10^{-16}$ |
| 2 | $D_2 = 0.3$, $D_4 = 0.05$ | 0.5 | 1 | 222.131 | 222.13 | $4.9 \times 10^{-6}$ |

## 3. Chirp and multi-line sources

Table S3 gives the corresponding checks for chirped pulses and multi-line sources, including asymmetric line sets. The line sets are given as (weight, offset $\Omega_j T$), and each line has width $V$. In all cases the mean output delay predicted by Eq. (17) also agrees with the simulation to at least five decimal places.

**Table S3. rms Width with Chirp and Multi-Line Sources**

| $m$ | Coefficients | $V$ | $C$ | Source lines | Eq. (17) | Numerical | Relative error |
|---|---|---|---|---|---|---|---|
| 1 | $D_2 = 0.4$, $D_3 = 0.2$ | 0.5 | -3 | 1 line | 40.5825 | 40.5825 | $< 10^{-16}$ |
| 2 | $D_2 = 0.3$, $D_3 = 0.1$ | 0 | -3 | 1 line | 262.686 | 262.686 | $6.5 \times 10^{-16}$ |
| 3 | $D_2 = -0.2$, $D_3 = 0.05$ | 0.5 | -2 | 1 line | 78.8856 | 78.8856 | $2.2 \times 10^{-15}$ |
| 2 | $D_3 = 0.1$ | 0 | -3 | 1 line | 257.212 | 257.212 | $1.3 \times 10^{-15}$ |
| 1 | $D_2 = 0.3$, $D_3 = 0.2$ | 0.3 | 0 | (0.5, -2); (0.5, 2) | 10.9785 | 10.9785 | $3.2 \times 10^{-16}$ |
| 2 | $D_2 = 0.3$, $D_3 = 0.1$ | 0.3 | 0 | (0.6, -1.5); (0.3, 0.5); (0.1, 3) | 10.5744 | 10.5744 | $1.7 \times 10^{-16}$ |
| 3 | $D_2 = 0.2$, $D_3 = 0.05$ | 0.5 | 1 | (0.7, 0); (0.3, 2.5) | 19.5728 | 19.5728 | $1.8 \times 10^{-16}$ |

## 4. Pulse pairs

The two-pulse results of Section 8 were checked in three ways. First, Eq. (30) was compared with the direct evaluation of Eq. (28) for Gaussian pulses separated by $\tau = 4T$, with $D = 0.5$ and 2, $V = 0$, 1 and 2, and $\theta = 0$, $\pi/2$ and $\pi$; the largest absolute difference in the normalized output power is $1.7 \times 10^{-13}$. Second, the energy, centroid and variance of the pulse pair were compared with those of the incoherent sum for $\theta = 0$ and $\pi$, with $D = 2$ or $B = 0.3$. For $m = 3$ the relative differences are below $10^{-15}$ with $\beta_2$ and $2 \times 10^{-9}$ with $\beta_3$, as expected from Eq. (29) and the input overlap of $10^{-29}$, whereas for Gaussian pulses, whose input overlap is 0.018, they reach 10%. Third, the midpoint visibility computed at $\xi = 16$ agrees with the limit of Eq. (32) to within $1.2 \times 10^{-3}$ for $m = 1$ and 3 and $V = 0.5$, 1 and 2.

## 5. Simulation window

**Table S4. Convergence of the Numerical rms Width with the Simulation Window for $m = 4$, $D = 0$, $B = 0.3$, $V = C = 0$**

| $L/T$ | $N$ | $\sigma^2/\sigma_0^2$ (numerical) | Relative error |
|---|---|---|---|
| 50 | $2^{16}$ | 67.758636 | $4.3 \times 10^{-1}$ |
| 200 | $2^{18}$ | 114.428671 | $3.9 \times 10^{-2}$ |
| 1000 | $2^{20}$ | 119.118716 | $4.5 \times 10^{-6}$ |
| 3000 | $2^{22}$ | 119.119248 | $3.0 \times 10^{-12}$ |

Equation (21) gives $\sigma^2/\sigma_0^2 = 119.119248$.

## 6. Critical dispersion coefficients

Table S5 lists the critical normalized coefficients used for the capacity curves of Section 8.6. The eye criterion was evaluated by bisection on $D_n$ with the ten periodic 100-bit sequences of Section 8.5, $\tau = 4T$, all bits in phase and $V = 0$, up to a relative precision of about $10^{-3}$. The rms criterion follows from Eq. (17).

**Table S5. Critical Normalized Dispersion Coefficients $D_{n,c}$ for a 1-dB Eye-Opening Penalty and for the rms Criterion $\sigma \leq T$**

| $n$ | $m = 1$: eye | $m = 1$: rms | $m = 2$: eye | $m = 2$: rms | $m = 3$: eye | $m = 3$: rms | $m = 4$: eye | $m = 4$: rms |
|---|---|---|---|---|---|---|---|---|
| 2 | 0.365 | 0.5 | 0.41 | 0.404 | 0.385 | 0.335 | 0.359 | 0.291 |
| 3 | 0.303 | 0.333 | 0.253 | 0.132 | 0.238 | 0.066 | 0.234 | 0.0407 |
| 4 | 0.0901 | 0.129 | 0.0619 | 0.0228 | 0.0498 | 0.00721 | 0.0486 | 0.0032 |
| 5 | 0.0668 | 0.0577 | 0.0331 | 0.00364 | 0.0216 | 0.000665 | 0.019 | 0.000204 |
| 6 | 0.0272 | 0.0217 | 0.0125 | 0.000501 | 0.0119 | 5.11e-05 | 0.00627 | 1.07e-05 |

## 7. Code

The Python code used to produce the figures and tables implements Eqs. (12), (13), (17) and (18) in 50-digit arithmetic, the ensemble-averaged propagation of Eq. (3), and the symbolic checks described above. It is available from the author.